\documentclass[journal]{IEEEtran}
\usepackage{cite}
\usepackage{mathtools}
\usepackage{gensymb}
\usepackage[caption=false,font=footnotesize]{subfig}
\usepackage[ruled,vlined]{algorithm2e}
\usepackage{amsmath,amssymb,amsfonts}
\usepackage{pifont}
\usepackage{textcomp}
\usepackage{tikz}
\usepackage{tikzscale}
\usepackage{circuitikz}
\usepackage[acronyms,nonumberlist]{glossaries}

\newacronym{BS}{BS}{base station}
\newacronym[plural=UEs,
            longplural={user equipment}]{UE}{UE}{user equipment}
\newacronym{UL}{UL}{uplink}
\newacronym{DL}{DL}{downlink}
\newacronym{PASS}{PASS}{pinching-antenna system}
\newacronym{PA}{PA}{pinching antenna}
\newacronym{LoS}{LoS}{line-of-sight}
\newacronym{NLoS}{NLoS}{non-line-of-sight}
\newacronym{AWGN}{AWGN}{additive white Gaussian noise}
\newacronym{SNR}{SNR}{signal-to-noise ratio}
\newacronym{SINR}{SINR}{signal-to-interference-plus-noise ratio}
\newacronym{SIR}{SIR}{signal-to-interference ratio}
\newacronym{CDF}{CDF}{cumulative distribution function}
\newacronym{PDF}{PDF}{probability density function}
\newacronym{RV}{RV}{random variable}
\newacronym{BM}{BM}{Brownian motion}
\newacronym{TA}{TA}{tangent approximation}
\newacronym{REM}{REM}{radio environment map}
\newacronym{NN}{NN}{nearest neighbor}
\newacronym{BI}{BI}{bilinear interpolation}
\newacronym{OMA}{OMA}{orthogonal multiple access}
\newacronym{NOMA}{NOMA}{non-orthogonal multiple access}
\newacronym{LSF}{LSF}{large-scale fading}
\newacronym{SSF}{SSF}{small-scale fading}
\newacronym{RIS}{RIS}{reconfigurable intelligent surface}
\newacronym{FAS}{FAS}{fluid antenna system}
\newacronym{SU}{SU}{single-user}
\newacronym{BER}{BER}{bit error rate}
\newacronym{LCX}{LCX}{leaky coaxial cable}
\newacronym{WG}{WG}{waveguide}
\newcommand{\tick}{\ding{51}}
\newcommand{\cross}{\ding{55}}

\SetKwBlock{Collect}{Sample collection phase}{}
\SetKwBlock{Average}{Spatial averaging phase}{}

\newtheorem{remark}{Remark}

\begin{document}

\title{Analysis and Antenna Position Selection for Pinching-Antenna Systems in Fading Channels}
\author{Amy~S.~Inwood,~\IEEEmembership{Member,~IEEE,} Peter~J.~Smith,~\IEEEmembership{Fellow,~IEEE,}
Hyundong~Shin,~\IEEEmembership{Fellow,~IEEE,} and~Michail~Matthaiou,~\IEEEmembership{Fellow,~IEEE}

\thanks{This work was supported by the U.K. Engineering and Physical Sciences Research Council (EPSRC) grant (EP/X04047X/2) for TITAN Telecoms Hub. The work of P. J. Smith was supported by the Marsden Fund Council from New Zealand Government funding, managed by Royal Society Te Apārangi. The work of H. Shin was supported by the National Research Foundation of Korea (NRF) grant funded by the Korean government (MSIT) under RS-2025-00556064 and by the MSIT (Ministry of Science and ICT), Korea, under the ITRC (Information Technology Research Center) support program (IITP-2025-RS-2021-II212046) supervised by the IITP (Institute
for Information \& Communications Technology Planning \& Evaluation). The work of M. Matthaiou was supported by the European Research Council (ERC) under the European Union’s Horizon 2020 Research and Innovation Programme (grant agreement No. 101001331).}

\thanks{A. S. Inwood is with the Centre for Wireless Innovation (CWI), Queen’s University Belfast, Belfast BT3 9DT, U.K. (e-mail: a.inwood@qub.ac.uk).}

\thanks{P. J. Smith is with the School of Mathematics and Statistics, Victoria University of Wellington, Wellington, New Zealand (e-mail: peter.smith@vuw.ac.nz).}

\thanks{H. Shin is with the Department of Electronics and Information Convergence Engineering, Kyung Hee University, Yongin-si, Gyeonggi-do 17104, Republic of Korea (e-mail: hshin@khu.ac.kr).}

\thanks{M. Matthaiou is with the Centre for Wireless Innovation (CWI), Queen’s University Belfast, BT3 9DT Belfast, U.K., and also with the
Department of Electronic Engineering, Kyung Hee University, Yongin-si, Gyeonggi-do 17104, Republic of Korea (e-mail: m.matthaiou@qub.ac.uk).}
}
 
\maketitle
\begin{abstract} Pinching-antenna systems (PASSs) are an emerging flexible antenna technology that enables the creation of line-of-sight (LoS) links by repositioning a radiating element along a dielectric waveguide (WG). Most existing studies of PASSs assume a purely LoS environment, despite the fact that many practical wireless environments exhibit scattering and spatially correlated shadow fading. In this paper, we present the first statistical analysis of PASS performance under Ricean fading with spatially correlated large-scale fading. Focusing on a fundamental single-user uplink PASS, we characterize the distribution and mean of the maximum achievable signal-to-noise ratio (SNR). For the classical exponentially correlated shadow fading scenario, we express the channel power process in terms of Brownian motion and derive highly accurate approximations for the maximum SNR distribution and mean. For a general class of smoothly correlated shadow fading processes, we derive an accurate expression for the SNR distribution by modeling the process in terms of the fade sojourn duration. To address the challenges of antenna position selection in fading environments, we propose radio environment map (REM)-based positioning strategies, including opportunistic random-look and grid-based constructions. Analytical results are validated by simulation, and the impacts of key system parameters, such as the Ricean $K$-factor, spatial correlation, WG length, and carrier frequency are investigated. The results provide insights into PASS operation beyond idealized LoS conditions and establish a general analytical framework for performance evaluation and antenna positioning in scattered environments.

\end{abstract}

\begin{IEEEkeywords}
Pinching-antenna systems, Rayleigh fading, Ricean fading, statistical analysis.
\end{IEEEkeywords}

\section{Introduction}
The demand for mobile connectivity is increasing rapidly with the rise of data-intensive applications, such as artificial intelligence, virtual reality, and high-definition streaming. One approach to meeting the capacity requirements of these networks is to increase the carrier frequency. However, higher frequencies create more challenging propagation conditions, such as increased path loss, attenuation, and shadowing.

Flexible antenna technologies \cite{ding_flexible_2025} are attracting increasing attention due to their ability to shape and manipulate the radio propagation environment. \Glspl{FAS}, which are reconfigurable on the wavelength scale, can enhance \gls{SSF} performance but cannot create new \gls{LoS} paths. In contrast, \glspl{RIS} can create \gls{LoS} links. However, this requires an additional propagation link, leading to amplified path losses.

\Glspl{PASS} are a new flexible antenna technology first proposed by NTT DOCOMO \cite{fukuda_pinching_2022}. \Glspl{PASS} consist of a simple dielectric particle (a \gls{PA}) placed on a dielectric \gls{WG} to radiate wireless signals. The dielectric \gls{WG}, spanning lengths of the order of meters, is fed by a \gls{BS} and routed along surfaces close to \glspl{UE}. Repositioning the \gls{PA} creates \gls{LoS} links without introducing an extra channel, while offering low cost and simple implementation. This has been practically demonstrated with a \gls{WG} constructed of polytetrafluoroethylene rods and pinching element formed by a small dielectric attached to a clothespin \cite{fukuda_pinching_2022}. As both components are common industrial materials and the \gls{WG} is passive, implementing a \gls{PASS} is likely several orders of magnitude cheaper than deploying an additional \gls{BS}.

While \glspl{PASS} can create \gls{LoS} links, most practical wireless environments exhibit scattering and shadowing effects in addition to any dominant \gls{LoS} component. In such settings, the channel experienced by a \gls{PA} is more accurately characterized by the joint presence of deterministic \gls{LoS} propagation and stochastic scattered components, together with spatially correlated large-scale fading. Understanding how these effects influence \gls{PASS} performance and \gls{PA} positioning is therefore essential for assessing their operation beyond the idealized assumption of a pure \gls{LoS} channel.

\subsection{Review of Related Literature}
As an emerging technology, \glspl{PASS} are attracting increasing research attention. However, the majority of existing studies focus on purely \gls{LoS} channels between the \gls{PA} and the \glspl{UE}. These studies fall into four main categories: those focusing on the \gls{SINR} and sum rate, outage probability, bit-level metrics and localization. Key works in each topic are summarized below.

The works in \cite{ding_flexible_2025, qian_pinching_2026,wang_antenna_2025,chen_sum_2026} focus on the \gls{SINR} and sum rate of \glspl{PASS} under the assumption of purely \gls{LoS} channels. As a foundational work in the \gls{PA} space, \cite{ding_flexible_2025} considered a range of system scenarios. The authors begin by deriving the ergodic sum rate for a \gls{PASS} consisting of a single \gls{PA} located on a single \gls{WG} attached to a single \gls{BS} serving a single \gls{UE}. For a system with multiple \glspl{PA} on a single \gls{WG}, they then derived the per-user rate upper bound in an \gls{OMA} system, and the ergodic rate for each \gls{UE} in a two-user \gls{NOMA} system. Finally, they proposed an algorithm to determine the optimal positions for each of two \glspl{PA} on their own \glspl{WG} serving two \glspl{UE}. 

In \cite{qian_pinching_2026}, the spectral efficiency of a system serving multiple \glspl{UE} was analyzed, and positioning strategies were developed for both distributed deployments (one \gls{PA} per \gls{WG}), and centralized deployments (multiple \glspl{PA} on a single \gls{WG}). Distributed deployments were found to be superior in the high-\gls{SNR} regime, while centralized deployments performed well in the low-\gls{SNR} regime.

The works in \cite{wang_antenna_2025} and \cite{chen_sum_2026} both investigated \glspl{PASS} involving a single \gls{WG} containing multiple \glspl{PA} serving multiple \glspl{UE}. In \cite{wang_antenna_2025}, a \gls{PA} position selection algorithm was proposed for a \gls{NOMA} \gls{PASS} that applied matching theory to maximize the sum rate. Meanwhile, \cite{chen_sum_2026} studied a multiple-access communication system in which \glspl{PA} were employed to generate multiple beamforming vectors within a single transmission period. The authors derived an upper bound on the achievable sum rate and show that the optimal sum rate performance is attained by alternating transmission among the \glspl{UE}, where for each transmission the PA placement is optimized to maximize the channel power gain of the selected \gls{UE}.

Another focus of \gls{LoS} \gls{PASS} work has been outage probability, due to the strong impact of \gls{LoS} availability on this metric. The works in \cite{tyrovolas_outage_2025,tyrovolas_performance_2026,ding_los_2025} investigate the outage probability of single-\gls{UE}, single-\gls{PA} \glspl{PASS}. In \cite{tyrovolas_outage_2025} and \cite{tyrovolas_performance_2026} the outage probability and the optimal \gls{PA} position to maximize the \gls{SNR} were derived. Additionally, \cite{tyrovolas_performance_2026} also derived the average rate for this system, and proposed a highly accurate scaling law to approximate the outage probability and average rate for a system with multiple \glspl{PA} on a single \gls{WG}. In \cite{ding_los_2025}, the impact of blockages between the \gls{PA} and the \gls{UE} was investigated. The authors derived the outage probability accounting for the possibility of a blockage in the \gls{LoS} link that renders service impossible. The \gls{LoS} blockage probability was modeled as a random variable that decays with the squared distance between the \gls{PA} and the \gls{UE}. Additionally, the ergodic rate was derived for a scenario where multiple \glspl{UE} are each served by one \gls{WG} with one \gls{PA}, again accounting for the possibility of blockages. The ergodic rate results showed that \gls{LoS} blockages can improve sum rate, due to their ability to reduce inter-user interference.

Focusing now on bit-level metrics, the work in \cite{alaaeldin_ber_2026} provided the first \gls{BER} analysis of a \gls{PA}-based \gls{NOMA} system. A single-antenna \gls{BS} served two \glspl{UE} via a single \gls{PA} on a single \gls{WG}. \gls{BER} expressions for both the \gls{UL} and \gls{DL} were derived, and these were then used to optimize the \gls{PA} location to minimize the average \gls{BER}.

Finally, the work in \cite{he_pinching_2026} studied the use of a \gls{PASS} for target localization. For a system involving \glspl{WG}, each with a single \gls{PA}, it derived the distribution of the \gls{SINR} at the target from a given \gls{PA} to evaluate localization performance and the Cram\'{e}r–Rao lower bound for received-signal-strength–based localization. The results demonstrated the superiority of \glspl{PASS} over conventional fixed-antenna systems for localization. 

These works derived important and insightful results for a wide range of \glspl{PASS} in purely \gls{LoS} environments. However, in many systems, scattered channel components may also exist, changing system analysis and adding additional complexities to \gls{PA} position selection through stochastic components.  

To date, limited works have considered the impact of \gls{NLoS} channels in \glspl{PASS}. In \cite{sun_stochastic_2026}, the outage probability was derived for a system with multiple \glspl{BS}, each connected to a \gls{WG} with a single \gls{PA} placed at one of a discrete set of locations. Each \gls{BS} serves multiple \glspl{UE}, while \glspl{UE} associated with other \glspl{BS} act as sources of interference. The channel between each \gls{PA} and \gls{UE} was modeled as either \gls{LoS} or \gls{NLoS}, determined using a rectangular Boolean modeling method. Although this work accounted for stochastic scattering, it assumed the channel to be solely \gls{LoS} or \gls{NLoS}, whereas in practice both components will likely coexist.

The work in \cite{wang_pinching_2025} studied a system in which multiple \glspl{UE} are each served by a \gls{WG} equipped with a single \gls{PA}. The channel between the \gls{PA} and each \gls{UE} comprised a deterministic \gls{NLoS} component determined by the scatterer locations, along with a \gls{LoS} component governed by a binary blockage indicator. The sum rate was maximized through the joint optimization of \gls{WG} assignment and \gls{PA} placement using matching theory. The results demonstrated that the \gls{LoS} component can be exploited to enhance the desired signal, while the \gls{NLoS} component can assist in mitigating inter-user interference. It was seen that an increased number of blockages improved the sum rate by more effectively suppressing inter-user interference. This work considered channels comprising both \gls{LoS} and \gls{NLoS} components, however, the results relied on the knowledge of scatterer location and size, leading to results tightly dependent on the specific geometry of the environment.

Additionally, \cite{wang_generalized_2026} and \cite{wang_leaky_2026} considered a system involving \glspl{LCX} acting as \glspl{WG}, multiple slots at discrete locations acting as \glspl{PA} and multiple \glspl{UE}. Here, the considered channels consist of both \gls{LoS} and \gls{NLoS} channel components. In these studies, the sum rate is maximized and the achievable data rate is obtained, assuming binary slot activation in \cite{wang_generalized_2026} and continuous slot aperture adjustment in \cite{wang_leaky_2026}. Again, results are reliant on knowledge of the specific environment geometry, thereby reducing generality.

\subsection{Contributions}
As discussed, limited prior work has examined the impact of scattered channel components in \glspl{PASS}, and none has considered fully  stochastic channels that include both a \gls{LoS} component and a stochastic scattered component with correlated fading. Motivated by this gap, we herein present the first study of a \gls{PASS} operating under Ricean fading, as well as the first considering spatially correlated shadow fading. This highly general framework accommodates \gls{PA} channels with arbitrary fading severity and provides insight into the transition from the idealized perfectly \gls{LoS} channels often assumed in the literature to channels with substantial scattered components. 

In this study, we consider the performance of a fundamental \gls{UL} \gls{PASS} in which a single-antenna \gls{UE} communicates with a \gls{PA} mounted on a \gls{WG} connected to a single \gls{BS} over a Ricean fading channel. The \gls{LSF} between the \gls{UE} and \gls{PA} incorporates spatially correlated shadowing, for which we consider both the exponential correlation model and a general class of smoothly correlated correlation models. Attenuation within the \gls{WG} is also considered.

The introduction of scattered channel components complicates the \gls{PA} position selection. In a \gls{SU} \gls{LoS} \gls{PASS}, the optimal \gls{PA} location depends solely on the distance between the \gls{PA} and the \gls{UE} and the loss within the \gls{WG}. Obtaining the optimal \gls{PA} location is more complicated in the presence of fading, due to the random nature of the channel. Thus, we investigate \glspl{REM} for \gls{PA} position selection. Originally introduced in the context of cognitive radio, \glspl{REM} capture and store spatially varying radio propagation information in a database that can inform system design \cite{zhao_network_2006, yilmaz_location_2015}. The slowly varying nature of correlated \gls{LSF} results in a relatively stable communication environment, making \glspl{REM} a well-suited tool for \gls{PA} position selection.

Therefore, we make the following novel contributions:
\begin{itemize}
    \item By expressing the exponentially correlated shadow fading process in terms of standard \gls{BM}, we derive highly accurate approximate expressions for the distribution and mean of the instantaneous maximum \gls{SNR}, taking into account both \gls{SSF} and \gls{LSF}.
    \item Similarly, for the general class of smoothly correlated shadow fading processes, we derive an accurate expression for the \gls{SNR} distribution by expressing the problem in terms of the fade sojourn duration of the process and exploiting the fact that this can be well modeled by an exponential distribution.
    \item We propose two \gls{REM} construction methods for \gls{PA} position selection in scattering environments: an opportunistic random-look approach and a grid-based approach. Using the resulting \glspl{REM}, we evaluate both nearest-neighbor and interpolation-based \gls{PA} selection strategies. For the grid-based method, we analyze the impact of localization errors on performance. We then compare the performance of these \gls{REM}-based strategies against random, optimal and nearest-to-\gls{UE} \gls{PA} positions. We also compare the performance of the considered continuous \gls{PASS} with a discrete counterpart.
\end{itemize}
Analytical results are verified by simulation, and the effects of key system parameters, such as Ricean $K$-factor, spatial correlation, \gls{PA} \gls{WG} size, carrier frequency, and shadow fading standard deviation are investigated. These contributions represent a comprehensive and cohesive treatment of the performance of a \gls{PASS} in a Ricean environment. Table \ref{tab:comparison} compares the contributions of our work to the state-of-the-art. 
\begin{table*}[t]
  \caption{The contributions of this work compared to existing literature on PASSs.}
    \centering
    \setlength{\tabcolsep}{5pt}
    \begin{tabular}{|c|c|c|c|c|c|c|c|c|c|c|c|c|c|c|}
    \hline
        \textbf{Contributions} & \textbf{This Work} & \cite{ding_flexible_2025} &\cite{qian_pinching_2026} & \cite{wang_antenna_2025} & \cite{chen_sum_2026} & \cite{tyrovolas_outage_2025} &  \cite{tyrovolas_performance_2026} &  \cite{ding_los_2025} & \cite{alaaeldin_ber_2026}&  \cite{he_pinching_2026}  &  \cite{sun_stochastic_2026} & \cite{wang_pinching_2025} & \cite{wang_generalized_2026} & \cite{wang_leaky_2026} \\
        \hline
        Pinching antenna systems & \tick & \tick & \tick & \tick & \tick & \tick & \tick & \tick & \tick & \tick & \tick & \tick & \tick & \tick \\ \hline
        Scattered channels & \tick & \cross & \cross & \cross & \cross & \cross & \cross & \cross & \cross & \cross & \tick & \tick & \tick & \tick \\
        \hline
        Ricean channels & \tick & \cross & \cross & \cross & \cross & \cross & \cross & \cross & \cross & \cross & \cross & \cross & \cross & \cross \\
        \hline
        Exponentially correlated fading & \tick & \cross & \cross & \cross & \cross & \cross & \cross &\cross & \cross & \cross & \cross & \cross & \cross & \cross\\
        \hline
        Smoothly correlated fading & \tick & \cross & \cross & \cross & \cross & \cross & \cross & \cross & \cross & \cross & \cross & \cross & \cross & \cross\\
        \hline
        SNR/rate analysis & \tick & \tick & \tick & \cross & \tick & \cross & \tick & \tick & \cross & \tick & \cross & \cross &\tick  & \tick \\
        \hline
        Outage probability analysis  & \tick & \cross & \cross & \cross & \cross & \tick & \tick & \tick & \cross & \cross & \tick & \cross & \cross & \cross \\
        \hline
        Pinching antenna position selection/beamforming  & \tick & \tick & \tick & \tick & \tick & \tick & \tick & \cross & \tick & \cross & \cross & \tick & \tick & \tick  \\
        \hline
        Radio environment maps & \tick & \cross & \cross & \cross & \cross & \cross & \cross & \cross & \cross & \cross & \cross & \cross &\cross &\cross \\
        \hline
    \end{tabular}
    \label{tab:comparison}
\end{table*}

\textit{Notation}: The set of complex numbers is denoted by $\mathbb{C}$; $\Re(z)$ represents the real component of complex number $z$; $\mathbb{E}[\cdot]$ and $\mathrm{Var}[\cdot]$ are the statistical expectation and variance;  $\mathcal{N}(\mu,\sigma^2)$ and $\mathcal{CN}(\mu,\sigma^2)$ denote the Gaussian and complex Gaussian distributions with mean $\mu$ and variance $\sigma^2$; $f_X(x)$ and $F_X(x)$ are the \gls{PDF} and \gls{CDF} of random variable $X$; $\varphi(\cdot)$ and $\Phi(\cdot)$ are the standard normal \gls{PDF} and \gls{CDF}; $\chi_2'^{2}(\cdot)$ is the non-central chi-squared distribution with 2 degrees of freedom; $P(A)$ denotes the probability of event $A$; $\mathrm{erf}(\cdot)$ is the error function;  $I_0(\cdot)$ is the zeroth-order modified Bessel function of the first kind; $Q_1(\cdot)$ is the first-order Marcum Q-function; $B(x)\sim\mathcal{N}(0,x)$ denotes standard \gls{BM}; $f'(x)=\frac{d}{dx}[f(x)]$ is the first derivative of $f(x)$ with respect to $x$; $(\cdot)^*$ denotes the complex conjugate; $\|\cdot\|_2$ denotes Euclidean distance; $\lfloor\cdot\rfloor$ is the floor operator; $\sup_{a<x<b} (\cdot)$ and $\max_{a<x<b}(\cdot)$ denote the supremum and maximum over $x \in (a,b)$, and $\arg\max_x(\cdot)$ and $\arg\min_x(\cdot)$ represent the values of $x$ that maximize and minimize the enclosed expression, respectively.

\section{System Model}

\subsection{Channel Model}
\label{sec:sysmod}
\begin{figure}[ht]
    \centering
    \resizebox{0.4\textwidth}{!}{\includegraphics{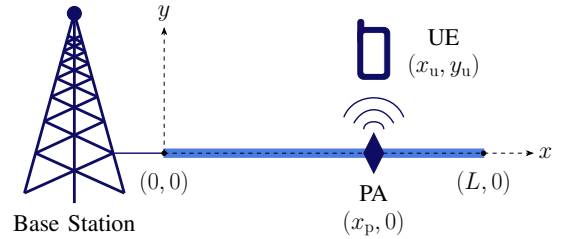}}
    \caption{The considered SU PASS architecture.}
    \label{fig:sys_model}
\end{figure}

Consider the \gls{PASS} shown in Fig.~\ref{fig:sys_model}, where a single-antenna \gls{BS} serves one single-antenna \gls{UE} via a single \gls{PA}. The \gls{WG} of length $L$ begins at $(x,y)=(0,0)$ and is aligned with the $x$-axis. The \gls{PA} can be activated at any position $\boldsymbol{\psi}_\mathrm{p}(x_\mathrm{p})=(x_\mathrm{p},0)$, $0\leq x_\mathrm{p}\leq L$, on the \gls{WG}. The \gls{UE} is located at $\boldsymbol{\psi}_\mathrm{u}=(x_\mathrm{u},y_\mathrm{u})$.

Let $h_\mathrm{p}(x_\mathrm{p})\in\mathbb{C}$ be the channel from the \gls{UE} to the \gls{PA}. We assume Ricean fading channels, such that
\begin{align}
    h_\mathrm{p}(x_\mathrm{p})\!&=\!\!\sqrt{\!\beta_\mathrm{LS}(x_\mathrm{p})}\Big(\!\sqrt{\!\tfrac{K}{K+1}}h_\mathrm{LoS}(x_\mathrm{p})\!+\!\sqrt{\!\tfrac{1}{K+1}}h_\mathrm{SC}(x_\mathrm{p})\!\Big)\!, \notag \\
    &=\!\!\sqrt{\!\beta_\mathrm{LS}(x_\mathrm{p})}\tilde{h}_\mathrm{p}(x_\mathrm{p}), \label{eq:htilde}
\end{align}
where $\beta_\mathrm{LS}(x_\mathrm{p})$ represents the \gls{LSF} loss between the \gls{UE} and \gls{PA}, $K$ is the Ricean $K$-factor, $h_\mathrm{LoS}(x_\mathrm{p})$ is the \gls{LoS} channel component and $h_\mathrm{SC}(x_\mathrm{p})$ is the scattered channel component. We adopt a Ricean channel model as a tractable alternative to highly detailed clustered ray-based models (e.g., \cite{3gpp_study_2020}), which, while accurate, are analytically complex. Prior studies show strong agreement between simplified scattered ray-based models and more detailed formulations for mmWave scenarios \cite{miller_analytical_2019}, supporting the aggregation of non-line-of-sight components into a Rayleigh term, with spatial consistency ensured via standard shadowing and path loss models. As this work is the first to consider Ricean fading channels for \glspl{PASS}, we begin by considering the fundamental scenario of a fixed $K$-factor, which allows the LoS component to be attenuated or removed entirely via $K$. The extension to a position-dependent $K$-factor is a valuable next step, and is discussed as a topic of future work in Sec. \ref{sec:FW}. The \gls{LoS} channel is modeled using the spherical wave channel model \cite{ding_flexible_2025,zhang_beam_2022, wang_antenna_2025}:
\begin{equation}  \label{sphr_ch_mod}
    h_\mathrm{LoS}(x_\mathrm{p}) = \mathrm{e}^{-j\frac{2\pi}{\lambda}d(x_\mathrm{p})},
\end{equation}
where $d(x_\mathrm{p}) = \left\|\boldsymbol{\psi}_\mathrm{u}-\boldsymbol{\psi}_\mathrm{p}(x_\mathrm{p})\right\|_2$ is the Euclidean distance between the \gls{UE} and \gls{PA}, $\lambda=\frac{c}{f_\mathrm{c}}$ is the wavelength, $c=3\times10^8$ m/s is the speed of light in a vacuum, and $f_\mathrm{c}$ is the carrier frequency. The scattered channel is modeled by Rayleigh fading, where $h_\mathrm{SC}(x_\mathrm{p})\sim \mathcal{CN}(0,1)$. The correlation between the scattered channels at the \gls{PA} positions $\boldsymbol{\psi}_\mathrm{p}(x_\mathrm{p})$ and $\boldsymbol{\psi}_\mathrm{p}(x_\mathrm{p}+\tau)$ depends on their separation, such that
\begin{equation}
    \mathbb{E}\left[h_\mathrm{SC}(x_\mathrm{p})h^*_\mathrm{SC}(x_\mathrm{p}+\tau)\right]= \begin{cases}
        \rho_\mathrm{SS}(\tau)& \quad\tau < d_\mathrm{SS},\\
        0 & \quad\tau \geq d_\mathrm{SS},
    \end{cases}
\end{equation}
where $\rho_\mathrm{SS}(\tau)$ represents any spatial correlation model and $d_\mathrm{SS}$ is the \gls{SSF} decorrelation distance. Applying the classic path loss and shadowing model \cite{goldsmith_wireless_2005} leads to a \gls{LSF} loss of
\begin{equation}
    \beta_\mathrm{LS}(x_\mathrm{p}) = A\,d(x_\mathrm{p})^{-\gamma}L(x_\mathrm{p}),
\end{equation}
where $A=\big(\frac{\lambda}{4\pi d_0}\big)^2$, $d_0=1$ m is the reference distance, $\gamma$ is the path loss exponent, $L(x_\mathrm{p})=10^{\sigma_\mathrm{SF}X(x_\mathrm{p})/10}$ is spatially correlated shadow fading driven by the correlated Gaussian process $X(x_\mathrm{p})\sim\mathcal{N}(0,1)$, and $\sigma_\mathrm{SF}$ is the shadow fading standard deviation. In this work, we consider exponentially correlated shadow fading \cite{gudmundson_correlation_1991}, where
\begin{equation}
    \label{eq:exp_corr}
    \rho_\mathrm{LS}(\tau)=\mathbb{E}[X(x_\mathrm{p})X(x_\mathrm{p}+\tau)]=\mathrm{e}^{-\alpha|\tau|},
\end{equation}
with $\alpha{=}1/d_\mathrm{LS}$ and $d_\mathrm{LS}$ denotes the \gls{LSF} decorrelation distance (the distance at which the correlation between shadow fading instances $X(x_\mathrm{p})$ and $X(x_\mathrm{p}+\tau)$ is $1/e$). We also consider a class of smoothly correlated shadow fading processes, where
\begin{equation}\label{eq:smooth_corr}
\rho_\mathrm{LS}(\tau)\!=\mathbb{E}[X(x_\mathrm{p})X(x_\mathrm{p}+\tau)]\sim 1 -a\tau^2 \quad \mathrm{as} \quad \tau \rightarrow 0, 
\end{equation}
for some $a>0$. Both types of correlation model are of interest. The classic exponential model has $\rho_\mathrm{LS}(\tau)\sim1-\alpha\tau$ for small $\tau$. This structure implies that $X(x_\mathrm{p})$ is a ``spiky" process and the derivative $\frac{d}{dx_\mathrm{p}}X(x_\mathrm{p})$ has infinite variance. Physically, this might be considered as a model of sharp discontinuities in the environment. In contrast, the smoothly correlated model has a finite variance for $\frac{d}{dx_\mathrm{p}}X(x_\mathrm{p})$, which ensures mean-square differentiability. This model leads to smoother variations across the shadow fading process.

We model the \gls{WG} loss as \cite{ding_flexible_2025}
\begin{equation}
    \beta_\mathrm{W}(x_\mathrm{p}) = 10^{-\frac{\kappa}{20}x_\mathrm{p}}\mathrm{e}^{-j\frac{2\pi}{\lambda_\mathrm{g}}x_\mathrm{p}},
\end{equation}
where $\kappa$ is the average attenuation factor along the dielectric \gls{WG} in dB/m, $\lambda_\mathrm{g}=\frac{\lambda}{n_\mathrm{eff}}$ is the guided wavelength, and $n_\mathrm{eff}$ is the effective refractive index of the dielectric \gls{WG}. Therefore, the channel between the \gls{UE} and the \gls{BS} via a \gls{PA} located at $\boldsymbol{\psi}_\mathrm{p}(x_\mathrm{p})$ is  
\begin{equation}
    h(x_\mathrm{p})= \beta_\mathrm{W}(x_\mathrm{p})h_\mathrm{p}(x_\mathrm{p}). 
\end{equation}

\subsection{Statistical Characterization of the SNR}
\label{sec:stat_char}
The received signal at the BS can be expressed as
\begin{equation}
    y=h(x_\mathrm{p})s+n,
\end{equation}
where $s$ is the transmitted signal, with $\mathbb{E}[|s|^2]=E_s$, and $n\sim\mathcal{CN}(0,\sigma^2)$ is the additive white Gaussian noise. Therefore, the optimal instantaneous \gls{SNR} is
\begin{align}
    \mathrm{SNR^{opt}}&=\frac{E_s}{\sigma^2}\sup_{0\leq x_\mathrm{p}\leq L}\left\{|h(x_\mathrm{p})|^2\right\}, \notag \\
    &=\sup_{0\leq x_\mathrm{p}\leq L}\left\{\beta(x_\mathrm{p})|\tilde{h}_\mathrm{p}(x_\mathrm{p})|^2\right\},
\end{align}
where $\beta(x_\mathrm{p})= \frac{E_s}{\sigma^2}|\beta_\mathrm
W(x_\mathrm{p})|^2\beta_\mathrm{LS}(x_\mathrm{p})$. In contrast, the long-term optimal \gls{SNR} is dependent only on \gls{LSF} and the loss within the \gls{WG}, and is
\begin{equation}
    \mathrm{SNR_{LT}^{opt}}= \sup_{0\leq x_\mathrm{p}\leq L}\left\{\beta(x_\mathrm{p})\right\}.
\end{equation}
The position at which $\mathrm{SNR_{LT}^{opt}}$ occurs is referred to as $x_{\mathrm{p,LT}}^{\mathrm{opt}}$. One effective strategy to identify suitable $x_{\mathrm{p,LT}}$ values is the use of an \gls{REM}, a database of information about a radio environment. The slowly varying nature of long-term \gls{SNR} ensures that measurements collected at a location remain valid over an extended period, making it a suitable candidate for a \gls{REM}. \Gls{REM} construction methods are detailed in Sec. \ref{sec:REM}. 

While identifying the global optimal \gls{PA} position is desirable, doing so reliably over a continuous \gls{WG} with a highly oscillatory instantaneous \gls{SNR} response would require prohibitive computational complexity. We instead  accurately approximate it by decomposing the problem into two stages. In the first stage, we identify $x_{\mathrm{p,LT}}^{\mathrm{opt}}$ to obtain $\mathrm{SNR_{LT}^{opt}}$. This is a well-established technique, analogous to designs based on statistical rather than instantaneous channel state information \cite{zheng_two_2025,hu_two_2025}. In the second stage, we search over $N$ independent \gls{SSF} samples in the vicinity of $x_{\mathrm{p,LT}}^{\mathrm{opt}}$. This principle is closely related to port selection in fluid antenna systems, where a set of closely-spaced candidate positions is evaluated to identify the one offering the best instantaneous channel realization. As in that setting, selecting points spaced sufficiently far apart to be approximately independent yields the greatest gain. Otherwise, samples will be correlated and thus more likely to take on similar values, reducing the likelihood of encountering a strong \gls{SSF} realization \cite{ramirez_new_2024}. For the same reason, extending the sampled region beyond a certain length yields diminishing returns, as a favorable \gls{SSF} realization has, with high probability, already been captured within it \cite{inwood_spatial_2026}. Therefore, we focus a \gls{SSF} search in the vicinity of $x_\mathrm{p,LT}^\mathrm{opt}$.

The two-stage approach achieves performance closest to $\mathrm{SNR^{opt}}$ when $\beta(x_\mathrm{p})$ and $|\tilde h_\mathrm{p}(x_\mathrm{p})|^2$ decorrelate over sufficiently different distances, i.e., $d_\mathrm{LS}\gg d_\mathrm{SS}$. In this case, the search incurs limited decay in $\beta(x_\mathrm{p})$ while capturing most of the achievable \gls{SSF} diversity gain. For typical system parameters, this requirement is comfortably satisfied.

As $\beta(x_\mathrm{p})$ and $\tilde{h}_\mathrm{p}(x_\mathrm{p})$ are independent, the \gls{SNR} obtained using this two-stage approach can be expressed as\footnote{Strictly speaking, this is an approximation, as $\mathrm{SNR_{LT}}$ will vary slightly over the search space. However, as the search space is in the order of wavelengths, this variation is assumed to be negligible.}
\begin{equation}
    \mathrm{SNR}_\mathrm{TS}=\mathrm{SNR_{LT}^{opt}}|\tilde{h}_\mathrm{p}^\mathrm{max}|^2, \label{eq:SNRopt}
\end{equation}
where 
\begin{equation}
    |\tilde{h}_\mathrm{p}^\mathrm{max}|^2=\max_{1\leq n\leq N}\left\{|\tilde{h}_\mathrm{p}(x_{\mathrm{p},n})|^2\right\},
\end{equation} 
is the maximum fading channel gain among the $N$ points sampled between $x_\mathrm{p,LT}^\mathrm{opt}-\Delta$ and $x_\mathrm{p,LT}^\mathrm{opt}+\Delta$, $\Delta=\delta(N-1)/2$, and $\delta>\lambda/2$ is the distance between independent samples.
The distribution of \eqref{eq:SNRopt} is
\begin{equation}
    P\left(\mathrm{SNR}_\mathrm{TS}\leq x\right)=P\Big(\mathrm{SNR_{LT}^{opt}}\leq \tfrac{x}{|\tilde{h}_\mathrm{p}^\mathrm{max}|^2}\Big).
\end{equation}
By the law of total probability,
\begin{align}
    P\left(\mathrm{SNR}_\mathrm{TS}\leq x\right)&=\mathbb{E}\Big[F_{\mathrm{SNR_{LT}^{opt}}}\Big(\tfrac{x}{|\tilde{h}_\mathrm{p}^\mathrm{max}|^2}\Big)\Big], \notag \\
    &= \int_0^\infty F_{\mathrm{SNR_{LT}^{opt}}}\left(\tfrac{x}{z}\right)f_{|\tilde{h}_\mathrm{p}^\mathrm{max}|^2}(z)\,dz, \label{eq:SNRmaxdist}
\end{align}
where $F_{\mathrm{SNR_{LT}^{opt}}}(\cdot)$ denotes the \gls{CDF} of the long-term optimal \gls{SNR}, and $f_{|\tilde{h}_\mathrm{p}^\mathrm{max}|^2}(\cdot)$ denotes the \gls{PDF} of the maximum fading channel gain over $N$ points. The distribution of the optimal long-term \gls{SNR} is dependent on the correlation model considered, and will be derived in following sections.

As the Ricean fading channels in \eqref{eq:htilde} are considered, the instantaneous channel power for a single sample position follows the scaled non-central chi-squared distribution with 2 degrees of freedom
\begin{equation}
    2(K+1)|\tilde{h}_\mathrm{p}(x_{\mathrm{p}})|^2\sim\chi_2'^{2}(2K),
\end{equation}
with the \gls{PDF} \cite[Eq. (2.21)]{simon_probability_2006}
\begin{equation}
    \!f_{|\tilde{h}_\mathrm{p}(x_\mathrm{p})|^2}(z)\!=\!(K\!+\!1)\mathrm{e}^{-K-(K+1)z}I_0\big(2\sqrt{K(K\!+\!1)z}\big),
\end{equation}
and the \gls{CDF} \cite[Eq. (2.22)]{simon_probability_2006}
\begin{equation}
    F_{|\tilde{h}_\mathrm{p}(x_\mathrm{p})|^2}(z)=1-Q_1\big(\sqrt{2K},\sqrt{2(K+1)z}\big).
\end{equation}
It is assumed that the $N$ fast fading channels considered are sampled at sufficiently large spacings to be uncorrelated. Therefore, using the order statistic result in \cite[Eq. (2.1.6)]{david_order_2003}, we obtain
\begin{equation}
    f_{|\tilde{h}_\mathrm{p}^\mathrm{max}|^2}(z)=Nf_{|\tilde{h}_\mathrm{p}(x_\mathrm{p})|^2}(z)\Big[F_{|\tilde{h}_\mathrm{p}(x_\mathrm{p})|^2}(z)\Big]^{N-1}. \label{eq:PDFhpmax}
\end{equation}

\subsection{Relationship Between the SNR and SIR}
The methods used in this work to investigate the \gls{SNR} can also be applied exactly to investigate the \gls{SIR} of a \gls{PASS} involving a single \gls{PA} serving a single \gls{UE} in the presence of an interfering \gls{UE}, broadening the applicability of the results. The long-term \gls{SIR} at point $x_\mathrm{p}\in[0,L]$ can be defined as
\begin{align}
    \mathrm{SIR}(x_\mathrm{p})\!=\!\frac{\beta_U(x_\mathrm{p})}{\beta_1(x_\mathrm{p})}&\!=\! \frac{Ad_{U}(x_\mathrm{p})^{-\gamma}10^{\sigma_\mathrm{SF}X_U(x_\mathrm{p})/10}}{Ad_{1}(x_\mathrm{p})^{-\gamma}10^{\sigma_\mathrm{SF}X_1(x_\mathrm{p})/10}}, \notag \\
    &\!=\! 10^{\sigma_\mathrm{SF}(X_U(x_\mathrm{p})-X_1(x_\mathrm{p}))/10}\!\left(\!\frac{d_U(x_\mathrm{p})}{d_1(x_\mathrm{p})}\!\right)^{\!\!-\gamma}\!,\notag \\ 
    &\!=\! 10^{\sigma_\mathrm{SF}X_c(x_\mathrm{p})/10}d_c(x_\mathrm{p})^{-\gamma},
\end{align}
where $d_c(x_\mathrm{p})=\frac{d_U(x_\mathrm{p})}{d_1(x_\mathrm{p})}$ is the ratio of the distance from the desired \gls{UE} to the \gls{PA} at position $x_\mathrm{p}$, $d_U(x_\mathrm{p})$, and the distance from the interferer to the \gls{PA} at position $x_\mathrm{p}$, $d_1(x_\mathrm{p})$, and $X_U(x_\mathrm{p})-X_1(x_\mathrm{p})\sim\mathcal{N}(0,2)$. For a simple model where $X_U(x_\mathrm{p})$ and $X_1(x_\mathrm{p})$ are independent, we obtain
\begin{align}
    \mathrm{Cov}[X_c(x_\mathrm{p}),X_c(x_\mathrm{p}+\tau)]&=\mathbb{E}[X_c(x_\mathrm{p})X_c(x_\mathrm{p}+\tau)], \notag \\
    &= \mathbb{E}[X_U(x_\mathrm{p})X_U(x_\mathrm{p}+\tau)] \notag \\ &\quad+\mathbb{E}[X_1(x_\mathrm{p})X_1(x_\mathrm{p}+\tau)], \notag \\ &=2\mathrm{e}^{-\alpha\tau}.
\end{align}
Hence, the correlation structure is the same as that described in Sec. \ref{sec:sysmod}, and the problem is the same as the \gls{SU} case involving no interference. Therefore, analysis undertaken for the \gls{SNR} is directly applicable to the \gls{SIR} for a single interferer.

\section{Analysis for Exponentially Correlated Large-Scale Fading}
\label{sec:expcorr_analysis}
In this section, we consider the shadow fading process $X(x_\mathrm{p})$ to be exponentially correlated, as in \eqref{eq:exp_corr}. We express this process in terms of standard \gls{BM} to facilitate the derivation of analytical expressions for the statistical properties of the \gls{SNR}. We derive a highly accurate approximation to the distribution of the optimal long-term \gls{SNR}, which can be combined with the results in \eqref{eq:SNRmaxdist} and \eqref{eq:PDFhpmax} to obtain an expression for the distribution of $\mathrm{SNR}_\mathrm{TS}$. The mean of $\mathrm{SNR}_\mathrm{TS}$ is then derived. 

\subsection{Optimal Long-Term SNR Distribution}
\label{subsec:LTSNR}
The distribution of the optimal long-term \gls{SNR} can be obtained considering its exceedance probability, as
\begin{equation}
    F_{\mathrm{SNR_{LT}^{opt}}}(x)=P(\mathrm{SNR_{LT}^{opt}}\leq x) = 1 - P(\mathrm{SNR_{LT}^{opt}}\geq x).
\end{equation}
The probability of $\mathrm{SNR_{LT}^{opt}}$ exceeding some threshold $x$ is equivalent to $\beta(x_\mathrm{p})$ exceeding $x$ anywhere between 0 and $L$. This itself can be split into two disjoint events, such that
\begin{equation}
    P(\mathrm{SNR_{LT}^{opt}}\geq x) = P_1 + P_2, \label{eq:exp_events}
\end{equation}
with
\begin{equation}
    P_1=P(\beta(0)>x),
\end{equation}
\begin{equation}
P_2= P(\beta(0)\leq x\,\cap\,\exists\, x_\mathrm{p}\in[0,L]:\beta(x_\mathrm{p})=x).
\end{equation}
The first event, $P_1$, is the probability that the process $\beta(x_\mathrm{p})$ is above the threshold when $x_\mathrm{p}=0$. Let
\begin{align}
    P_1&=1-P(\beta(0)\leq x) ,\notag \\ &=1-\!P\!\left(\tfrac{E_s}{\sigma^2}|\beta_\mathrm{W}(0)|^2A\,10^{\sigma_\mathrm{SF}X(0)/10}d(0)^{-\gamma}\!\leq\! x\right)\!. \label{eq:Fbeta0_original} 
\end{align}
Rewriting \eqref{eq:Fbeta0_original} in terms of $X(0)$,
\begin{align}
    P_1&=1-P\left(X(0)\leq\tfrac{10}{\sigma_\mathrm{SF}}\log_{10}\!\left(\tfrac{\sigma^2x}{E_sA|\beta_\mathrm{W}(0)|^2}\,d(0)^{\gamma}\right)\!\right)\!, \notag \\
    &=1-\Phi\left(\tfrac{10}{\sigma_\mathrm{SF}}\log_{10}\!\left(\tfrac{\sigma^2x}{E_sA|\beta_\mathrm{W}(0)|^2}\,d(0)^{\gamma}\right)\!\right)\!. \label{eq:P1_final}
\end{align}
The second event in \eqref{eq:exp_events}, $P_2$, is the probability that $\beta(x_\mathrm{p})$ is below the threshold when $x_\mathrm{p}=0$, but crosses the threshold at some other point between 0 and $L$. Rewriting $P_2$ in terms of $X(x_\mathrm{p})$,
\begin{equation}
    P_2\! =\! P\!\left(\!X(0)\!\leq \!y(0,x)\! \cap\! \exists x_\mathrm{p}\!\!\in\![0,L]\!:\!X(x_\mathrm{p})\!=\!y(x_\mathrm{p},x)\!\right)\!, \label{eq:event2X}
\end{equation}
where $y(x_\mathrm{p},x)=\frac{10}{\sigma_\mathrm{SF}\log(10)}(\log(x)+\Psi(x_\mathrm{p}))$ and $\Psi(x_\mathrm{p})=\log\!\big(\frac{\sigma^2d(x_\mathrm{p})^\gamma}{E_sA|\beta_\mathrm{W}(x_\mathrm{p})|^2}\big)$. As $X(x_\mathrm{p})$ is exponentially correlated, it is a stationary Ornstein-Uhlenbeck process \cite{pavliotis_stochastic_2015}, and thus is Markovian. Since $X(x_\mathrm{p})$ is also Gaussian, it can be expressed as a transformation of standard \gls{BM} \cite{pavliotis_stochastic_2015}, such that
\begin{equation}
    X(x_\mathrm{p})=\mathrm{e}^{-\alpha x_\mathrm{p}}B\left(\mathrm{e}^{2\alpha x_\mathrm{p}}\right), \label{eq:BM_transform}
\end{equation}
where $B(t)\sim\mathcal{N}(0,t)$ denotes standard \gls{BM}. This closed-form representation is not possible for non-Markovian correlation functions (e.g. Gaussian, Jakes or sinc) as these processes require knowledge of previous states to reproduce the covariance exactly. Applying the transformation in \eqref{eq:BM_transform}, $P_2$ becomes
\begin{equation}
    P_2\!=\!P\!\left(\!B(1)\!\leq\! y(0,x) \!\cap\!\exists x_\mathrm{p}\!\!\in\![0,L]\!:\! B\big(\mathrm{e}^{2\alpha x_\mathrm{p}}\!\big)\!\!=\!y(x_\mathrm{p},x)\mathrm{e}^{\alpha x_\mathrm{p}}\!\right)\!\!.\! \notag
\end{equation}
By the law of total probability and conditioning on $B(1)$,
\begin{equation}
    P_2=\int_{-\infty}^{y(0,x)}\!f(u)P_c(u,x)du, \label{eq:P2_og}
\end{equation}
where
\begin{equation}
    P_c(u,x)\!=\!P\!\left(\exists\,x_\mathrm{p}\!\!\in\!\![0,\!L]\!\!:\!\!B(\!\mathrm{e}^{2\alpha x_\mathrm{p}}\!)\!\!=\!y(x_\mathrm{p},x)\mathrm{e}^{\alpha x_\mathrm{p}}\!\!\mid\!\!B(1)\!=\! u\right)\!,\! \notag
\end{equation}
and $f(u)=\mathrm{e}^{-u^2/2}/\sqrt{2\pi}$ is the \gls{PDF} of $B(1)$. By the independent and stationary increment properties of \gls{BM}, $B(\mathrm{e}^{2\alpha x_\mathrm{p}})=u+B(\mathrm{e}^{2\alpha x_\mathrm{p}}-1)$. Therefore,
\begin{align}
    \!P_c(u,x)&\!=\! P\!\left(\exists\,x_\mathrm{p}\!\in \![0,L]\!:\!B\big(\mathrm{e}^{2\alpha x_\mathrm{p}}\!-\!1\big)\!=\!y(x_\mathrm{p},x)\mathrm{e}^{\alpha x_\mathrm{p}}\!-\!u\!\right)\!, \notag \\
    &\!=\!P\!\left(\exists\,s\in (0,\mathrm{e}^{2\alpha L }-1]:B(s)=\gamma_u(s,x)\right)\!,
\end{align}
where $s=\mathrm{e}^{2\alpha x_\mathrm{p}}-1\in (0, \mathrm{e}^{2\alpha L}-1]$ and $\gamma_u(s,x)=g(s,x)-u$, with
\begin{align}
   g(s,x)&=y\left(\Omega(s)\right)\sqrt{s+1}, \notag \\
   &= \tfrac{10\sqrt{s+1}}{\sigma_\mathrm{SF}\log(10)}\log\!\left(\tfrac{\sigma^2x\,d\left(\Omega(s)\right)^\gamma}{E_sA\,|\beta_\mathrm{W}\left(\Omega(s)\right)|^2}\right),
\end{align}
and $\Omega(s)=\frac{\log(s+1)}{2\alpha}$. Expressing the problem in terms of $s$ recasts the conditional probability $P_c(u,x)$ as the probability that a standard \gls{BM}, starting from zero at $s=0$, intersects the $u$-dependent curved boundary $\gamma_u(s,x)$ over a finite interval, reducing the problem to a first-hitting-time evaluation. Proposed in \cite{ferebee_tangent_1982}, the \gls{TA} provides an approximate expression for $P_c(u,x)$, where
\begin{multline}
    P_c(u,x)\approx\int_0^{\mathrm{e}^{2\alpha L}-1}\tfrac{1}{\sqrt{2\pi s}}\exp\left(-\tfrac{(g(s,x)-u)^2}{2s}\right) \\ \times \left(\tfrac{g(s,x)-u}{s}-g'(s,x)\right)ds,\label{eq_Pcint}
\end{multline}
and, derived in Appendix A, the derivative of $g(s,x)$ with respect to $s$ is
\begin{multline}
    g'(s,x)\! = \tfrac{5}{\sigma_\mathrm{SF}\sqrt{s+1}}\Big[\tfrac{\kappa}{10\alpha}+\tfrac{1}{\log(10)} \\ \times\left(\tfrac{\gamma\left(\Omega(s)-x_\mathrm{u}\right)}{\alpha d^2(\Omega(s))}+\log\!\left(\!\tfrac{\sigma^2x\,d\left(\Omega(s)\right)^\gamma}{E_sA|\beta_\mathrm{W}\!\left(\Omega(s)\right)|^2}\right)\!\right)\!\!\Big]. \label{eq:gdash}
\end{multline}
\begin{remark}
    The accuracy of the \gls{TA} depends on how well the threshold, $\gamma_u(s,x)$, can be approximated by a line. The dominant $s$ term is $d(s)^\gamma$, meaning accuracy is highest for shorter \glspl{WG}, \glspl{UE} located near the end of the \gls{WG}, and low path loss exponents. A possible topic for future work is the application of the \gls{TA} in a piecewise manner to highly non-linear boundaries.
\end{remark}
Combining \eqref{eq:P2_og} with \eqref{eq_Pcint} gives
\begin{multline}
    P_2\approx\tfrac{1}{2\pi}\int_{-\infty}^{y(0,x)}\!\int_0^{\mathrm{e}^{2\alpha L}-1}\exp\left(-\tfrac{(g(s,x)-u)^2}{2s}-\tfrac{u^2}{2}\right) \\ \times\tfrac{1}{\sqrt{s}}\left(\tfrac{g(s,x)-u}{s}-g'(s,x)\right)dsdu. \label{eq:P2_doubleint}
\end{multline}
The integral over $u$ in \eqref{eq:P2_doubleint} can be evaluated analytically and is derived in Appendix B, giving \eqref{eq:P2_approx} at the top of the next page. 
\begin{figure*}[!t]
\begin{multline}
     P_2\approx\int_0^{\mathrm{e}^{2\alpha L}\!-\!1} \!\!\sqrt{\tfrac{1}{8\pi(s+1)}}\!\left(\!\tfrac{g(s,x)}{s}\!-\!g'(s,x)\!\right)\exp\!\left(\tfrac{g^2(s,x)}{2s}\!\!\left(\!\tfrac{1}{s+1}\!-\!1\right)\!\right)\!\!\left(\!1\!+\!\mathrm{erf}\!\left(\!\sqrt{\tfrac{s+1}{2s}}y(0,x)-\tfrac{g(s,x)}{\sqrt{2s(s+1)}}\right)\!\!\right)\!ds \\ -\!\!\! \int_0^{\mathrm{e}^{2\alpha L}\!-\!1}\!\tfrac{\mathrm{exp}\big(\!-\frac{g^2(s,x)}{2(s+1)}\!\big)}{2\pi s\sqrt{s}}\!\bigg[\!\sqrt{\!\tfrac{\pi s}{2(s+1)}}\tfrac{g(s,x)}{s+1}\!\!\left(\!\!1\!+\!\mathrm{erf}\!\left(\!\!\sqrt{\tfrac{s+1}{2s}}y(0,x)\!-\!\tfrac{g(s,x)}{\sqrt{2s(s+1)}}\!\!\right)\!\!\right)\!\!-\!\tfrac{s\exp\!\big(\!-\frac{s+1}{2s}\!\big(y(0,x)-\frac{g(s,x)}{s+1}\!\big)^{\!2}\big)}{s+1}\!\bigg]\! ds.\! \label{eq:P2_approx}
\end{multline}
\hrulefill
\end{figure*}
Therefore, a highly accurate approximation to  $F_\mathrm{SNR_{LT}^{opt}}(x)$ can be expressed in terms of a normal \gls{CDF} (see \eqref{eq:P1_final}) and a single integral over $x_\mathrm{p}$ (see \eqref{eq:P2_approx}). The integrand in \eqref{eq:P2_approx} is smooth and bounded, so can be computed quickly and efficiently, making the derived distribution accurate and practical.

\subsection{Mean SNR}
\label{sec:expcorr_analysis_mean}
Taking the expectation of $\mathrm{SNR}_\mathrm{TS}$ in \eqref{eq:SNRopt} and using the fact that the long-term \gls{SNR} and fading channels are independent, 
\begin{equation}
    \mathbb{E}\left[\mathrm{SNR}_\mathrm{TS}\right]=\mathbb{E}\left[\mathrm{SNR_{LT}^{opt}}\right]\mathbb{E}\left[|\tilde{h}_\mathrm{p}^\mathrm{max}|^2\right]. \label{eq:ESNRmax}
\end{equation}
The above terms will be derived in the following subsections.

\subsubsection{Mean Optimal Long-Term SNR}
From \cite[Eq. (5-53)]{papoulis_probability_2002}, as $\mathrm{SNR_{LT}^{opt}}\geq0$, 
\begin{equation}
    \mathbb{E}\left[\mathrm{SNR_{LT}^{opt}}\right]=\int^\infty_0P\left(\mathrm{SNR_{LT}^{opt}}>x\right)dx.
\end{equation}
Therefore, using the results derived for the optimal long-term \gls{SNR} distribution in Sec. \ref{subsec:LTSNR},
\begin{equation}
    \mathbb{E}\left[\mathrm{SNR_{LT}^{opt}}\right]=T_1+T_2-T_3, \label{eq:ELTSNRTs}
\end{equation}
where
\begin{equation}
    T_1=\int_0^\infty\left[1-\Phi\left(\tfrac{10}{\sigma_\mathrm{SF}}\log_{10}\!\left(\tfrac{\sigma^2x}{E_sA|\beta_\mathrm{W}(0)|^2}\,d(0)^{\gamma}\right)\!\right)\!\right]dx,
\end{equation}
\begin{multline}
    \!\!\!T_2=\!\!\int_0^{\mathrm{e}^{2\alpha L}\!-\!1} \!\!\!\int_0^\infty\!\!\sqrt{\!\tfrac{1}{8\pi(s+1)}}\!\left(\!\tfrac{g(s,x)}{s}\!-\!g'\!(s,x)\!\right)\!\exp\!\left(\!-\tfrac{g^2(s,x)}{2(s+1)}\!\right) \\ \times\left(\!1\!+\!\mathrm{erf}\!\left(\!\sqrt{\tfrac{s+1}{2s}}y(0,x)-\tfrac{g(s,x)}{\sqrt{2s(s+1)}}\right)\!\!\right)\!dxds, \label{eq:T2int}
\end{multline}
and
\begin{multline}
    \!\!\!T_3\!=\!\!\! \int_0^{\mathrm{e}^{2\alpha L}\!-\!1}\!\!\!\!\int_0^\infty\!\!\tfrac{\mathrm{exp}\!\big(\!-\frac{g^2\!(s,x)}{2(s+1)}\!\big)}{2\pi s\sqrt{s}}\!\bigg[\!\sqrt{\!\tfrac{\pi s}{2(s+1)}}\tfrac{g(s,x)}{s+1}\!\Big(\!1+\mathrm{erf}\Big(\!\!\sqrt{\!\tfrac{s+1}{2s}\!\!} \\ \!\times \!y(0,x)\!-\!\tfrac{g(s,x)}{\sqrt{2s(s+1)}}\!\Big)\!\!\Big)\!\!-\!\tfrac{s\exp\!\big(\!-\!\frac{s+1}{2s}\!\left(\!y(0,x)-\!\frac{g(s,x)}{s+1}\!\big)\!^{2}\!\right)}{s+1}\!\bigg]\! dxds.\!\! \label{eq:T3int}
\end{multline}
The evaluation of the integral in $T_1$ is straightforward, as 
\begin{align}
    T_1&=\mathbb{E}[\beta(0)]=\tfrac{E_s}{\sigma^2}|\beta_\mathrm{W}(0)|^2Ad(0)^{-\gamma}\mathbb{E}[L(0)], \notag \\
    &=\tfrac{E_s}{\sigma^2}|\beta_\mathrm{W}(0)|^2Ad(0)^{-\gamma}\exp\Big(\tfrac{1}{2}\big(\tfrac{\log(10)\sigma_\mathrm{SF}}{10}\big)^{2}\Big), \label{eq:T1}
\end{align}
since $L(0)$ is a log-normal variable. The integrals over $x$ in $T_2$ and $T_3$ are solved in Appendices C and D respectively, giving
\eqref{eq:T2} and \eqref{eq:T3} at the top of the following page, with 
\begin{figure*}[!t]
\vspace{-0.2in}
\begin{multline}
    T_2= \int_0^{\mathrm{e}^{2\alpha L}-1}\tfrac{\sigma_\mathrm{SF}\log(10)}{20(s+1)}\exp\left(\tfrac{\sigma_\mathrm{SF}^2\log(10)^2}{200}-\Psi(\Omega(s))\right)\Big[\Big(\tfrac{\sigma_\mathrm{SF}\log(10)(s+2)}{20s} \\ - \tfrac{1}{\alpha\sigma_\mathrm{SF}}\left(\tfrac{\kappa}{2}+\tfrac{5\gamma(\Omega(s)-x_\mathrm{u})}{\log(10)d(\Omega(s))^2}\right)\!\!\Big)(1-\mathrm{erf}(\Lambda))+\tfrac{(s+2)(\sqrt{s+1}-1)}{2s\sqrt{\pi(s+1-\sqrt{s+1})}}\exp(-\Lambda^2)\Big]ds.
    \label{eq:T2}
\end{multline}
\vspace{-0.1in}
\begin{multline}
    T_3\!= \!\!\!\int_0^{\mathrm{e}^{2\alpha L}-1}\!\!\tfrac{\sigma_\mathrm{SF}\log(10)}{20s(s+1)}\!\bigg[\!\exp\!\left(\!\tfrac{\sigma_\mathrm{SF}^2\log(10)^2}{200}\!-\!\Psi(\Omega(s))\!\right)\!\!\bigg[\!\sqrt{\!\tfrac{\sqrt{s+1}-1}{\pi\sqrt{s+1}}}\!\exp\!\left(-\Lambda^2\right)+\tfrac{\sigma_\mathrm{SF}\log(10)}{10}(1\!-\mathrm{erf}(\Lambda)\!)\!\bigg]\! -\!\tfrac{s}{\sqrt{\pi(s+1-\sqrt{s+1})}}\\ \times\exp\!\left(\!-\tfrac{\sqrt{s+1}(100(s+1)+5(1-\sqrt{s+1}))}{2s\sigma_\mathrm{SF}^2\log(10)^2}\!\left(\Psi(\Omega(s))\!-\!\Psi(0)\right)^2\!-\!\tfrac{1}{2}\!\left(\!\Psi(\Omega(s)\!)\!+\!\Psi(0)\!+\!\tfrac{s\sigma_\mathrm{SF}^2\log(10)^2}{200\sqrt{s+1}(1-\sqrt{s+1})}\!\right)\!\!\right)\!\!\bigg]\!. \label{eq:T3}
\end{multline}
\vspace{-0.1in}
\hrulefill
\end{figure*}
\begin{multline}
    \Lambda=\tfrac{1}{\sqrt{s+(\sqrt{s+1}-1)^2}}\Big(\tfrac{\sqrt{50(s+1)}}{\sigma_\mathrm{SF}\log(10)} \\ \times\left(\Psi(\Omega(s))-\Psi(0)\right)-\tfrac{\sigma_\mathrm{SF}\log(10)(\sqrt{s+1}-1)}{10\sqrt{2}}\Big). \label{eq:Lambda}
\end{multline}
Combining these and \eqref{eq:T1} with \eqref{eq:ELTSNRTs} gives the mean optimal long-term \gls{SNR} expression.

\subsubsection{Mean SSF Maximum}
If we again assume that $N$ fading channels are sampled at sufficient distances to be uncorrelated, the derived PDF in \eqref{eq:PDFhpmax} can be used to ascertain the mean, such that
\begin{equation}
    \mathbb{E}\left[|\tilde{h}_\mathrm{p}^\mathrm{max}|^2\right]=\int_0^\infty z\,f_{|\tilde{h}_\mathrm{p}^\mathrm{max}|^2}(z)dz. \label{eq:Eabshpmax2}
\end{equation}
Combining \eqref{eq:ELTSNRTs} (comprising of \eqref{eq:T1} - \eqref{eq:T3}) and \eqref{eq:Eabshpmax2} (using \eqref{eq:PDFhpmax}) as in \eqref{eq:ESNRmax} gives an accurate approximation to $\mathbb{E}[\mathrm{SNR_{TS}}]$.

\section{Analysis for Smoothly Correlated Large-Scale Fading}
\label{sec:smoothcorr_analysis}
In this section, we consider the shadow fading process $X(x_\mathrm{p})$ to be smoothly correlated, as in \eqref{eq:smooth_corr}. An exact expression or highly accurate approximation to $F_\mathrm{SNR_{LT}^{opt}}(x/z)$ for a general $x$ is unknown under this correlation model. As for the exponentially correlated scenario, we propose an accurate approximation to the distribution of the optimal long-term \gls{SNR}. To do so, we take an approach similar to that in \cite{siriwardana_analysis_2026}. The event that the optimal long-term \gls{SNR} falls below a threshold is equivalent to the \gls{SNR} process starting below the threshold at $x_\mathrm{p}=0$ and remaining below it across the entire \gls{WG}. The distribution can therefore be expressed in terms of the average fade duration of the process. In \cite{siriwardana_analysis_2026}, the authors exploited the fact that the fade sojourn duration of a smooth fading process is well modeled by an exponential distribution with mean equal to the average fade duration. To apply this here, we first derive the expected number of crossings across the \gls{WG}, which is then used to obtain the corresponding average fade duration required for the exponential-based approximation. The derived long-term distribution can then be combined with \eqref{eq:SNRmaxdist} and \eqref{eq:PDFhpmax} to accurately approximate the overall distribution.



\subsection{Optimal Long-Term SNR Distribution}

First, let us express $\mathrm{SNR_{LT}^{opt}}$ in $\mathrm{dB}$ as
\begin{equation}
    10\log_{10}(\mathrm{SNR_{LT}^{opt}})=\sup_{0\leq x_\mathrm{p}\leq L}\left\{\mu(x_\mathrm{p})+\sigma_\mathrm{SF}X(x_\mathrm{p})\right\},
\end{equation}
where $\mu(x_\mathrm{p}){=}10\log_{10}(\frac{E_s}{\sigma^2}A|\beta_\mathrm{W}(x_\mathrm{p})|^2){-}10\gamma\log_{10}(d(x_\mathrm{p}))$.
Thus, the distribution of $\mathrm{SNR_{LT}^{opt}}$ in $\mathrm{dB}$ can be defined as
\begin{align}
    P_{\mathrm{SC}}&=P(10\log_{10}(\mathrm{SNR_{LT}^{opt}})\leq x), \notag \\
    &=P(X(x_\mathrm{p})\leq \upsilon(x_\mathrm{p})),
\end{align}
where $x$ represents the threshold in $\mathrm{dB}$ and 
\begin{align}
    \upsilon(x_\mathrm{p})&=\tfrac{1}{\sigma_\mathrm{SF}}(x-\mu(x_\mathrm{p})), \notag \\
    &=\varrho+b\log\left((x_\mathrm{u}-x_\mathrm{p})^2+y_\mathrm{u}^2\right)+cx_\mathrm{p},
\end{align} 
where $\varrho=\frac{1}{\sigma_\mathrm{SF}}(x-\frac{10}{\log(10)}\log(AE_s/\sigma^2))$, $b=\frac{5\gamma}{\sigma_\mathrm{SF}\log(10)}$ and $c=\frac{\kappa}{\sigma_\mathrm{SF}}$ The event $P_\mathrm{SC}$ is equivalent to $10\log_{10}(\mathrm{SNR_{LT}})$ being below $x$ when $x_\mathrm{p}=0$, and remaining below when $0<x_\mathrm{p}\leq L$, which can be expressed mathematically as
\begin{align}
    P_\mathrm{SC}&=P(X(0)\leq\upsilon(0),N_\upsilon(L)=0), \notag \\
    &= \Phi(\upsilon(0))P(N_\upsilon(L)=0\mid X(0)\leq\upsilon(0)), \notag \\
    &=\Phi(\upsilon(0))P(D_\mathrm{FS}>L),
\end{align}
where $N_\upsilon(L)$ is the number of upcrossings of $\upsilon(x_\mathrm{p})$ by $X(x_\mathrm{p})$ in $[0,L]$ and $D_\mathrm{FS}$ is the fade sojourn distance. As in \cite{siriwardana_analysis_2026}, we can approximate $D_\mathrm{FS}$ with an exponential with a mean equivalent to the average fade duration, $\mathrm{AFD}_\upsilon$, such that
\begin{equation}
    D_\mathrm{FS}\sim\mathrm{Exp}\left(\frac{1}{\mathrm{AFD}_\upsilon}\right),
\end{equation}
and $\mathbb{E}[D_\mathrm{FS}]=\mathrm{AFD}_\upsilon$. Therefore,
\begin{equation}
    P_\mathrm{SC}\approx\Phi(\upsilon(0))\exp\left(-\frac{L}{\mathrm{AFD}_\upsilon}\right).
\end{equation}
Applying the standard identity to express the average fade duration in terms of the level crossing rate, $\mathrm{LCR}_\upsilon$, \cite{zlatanov_level_2008}
\begin{align}
    P_\mathrm{SC}&\approx\Phi(\upsilon(0))\exp\left(-\frac{L\times\mathrm{LCR}_\upsilon}{\Phi(\upsilon(0))}\right), \notag \\
    &=\Phi(\upsilon(0))\exp\left(-\frac{\mathbb{E}[N_\upsilon(L)]}{\Phi(\upsilon(0))}\right). \label{eq:PSC}
\end{align}
Applying the result in \cite[Eq. (12)]{leadbetter_crossings_1965} for the mean number of upcrossings across a curved threshold, we obtain
\begin{multline}
    \mathbb{E}[N_\upsilon(L)]=\!\!\int_0^L\!\varphi(\upsilon(x_\mathrm{p}))\bigg[2\sigma_2\,\varphi\left(\frac{\upsilon'(x_\mathrm{p})}{\sigma_2}\right)  \\ +\upsilon'(x_\mathrm{p})\left(2\Phi\left(\frac{\upsilon'(x_\mathrm{p})}{\sigma_2}\right)-1\right)\!\bigg]dx_\mathrm{p}, \label{eq:numcross}
\end{multline}
where 
\begin{equation}
    \upsilon'(x_\mathrm{p})=\frac{2b(x_\mathrm{p}-x_\mathrm{u})}{(x_\mathrm{u}-x_\mathrm{p})^2+y_\mathrm{u}^2}+c,
\end{equation}
and $\varphi(\cdot)$ is the standard normal \gls{PDF}, and, from \eqref{eq:smooth_corr}$, \sigma_2^2=\mathrm{Var}[X'(t)]=2a$. Any smooth correlation function can be used to obtain $\sigma_2$. The full expression for $\mathbb{E}[N_\upsilon(L)]$ in \eqref{eq:ENvL} at the top of the following page comprises of two integrals, both of which allow simple and efficient numerical integration. Therefore, combining \eqref{eq:PSC} and \eqref{eq:ENvL} gives an accurate approximation to the full \gls{SNR} distribution under smoothly correlated shadow fading.

\begin{figure*}[!t]
\begin{multline}
    \mathbb{E}[N_\upsilon(L)]=\frac{1}{\sqrt{2\pi}}\int_0^L\!\exp\!\left(\!-\frac{(\varrho+b\log((x_\mathrm{u}-x_\mathrm{p})^2+y_\mathrm{u}^2)+cx_\mathrm{p})^2}{2}\!\right)\!\Bigg[2\sqrt{\frac{a}{\pi}}\exp\!\left(-\frac{1}{4a}\!\left(\!\frac{2b(x_\mathrm{p}-x_\mathrm{u})}{(x_\mathrm{u}-x_\mathrm{p})^2+y_\mathrm{u}^2}\!+\!c\!\right)^{\!2}\right)\! \\ +\! \left(\!\frac{2b(x_\mathrm{p}{-}x_\mathrm{u})}{(x_\mathrm{u}{-}x_\mathrm{p})^2+y_\mathrm{u}^2}+c\!\right)\!\mathrm{erf}\!\left(\!\frac{1}{2\sqrt{a}}\!\left(\!\frac{2b(x_\mathrm{p}-x_\mathrm{u})}{(x_\mathrm{u}-x_\mathrm{p})^2+y_\mathrm{u}^2}\!+c\!\right)\!\!\right)\!\Bigg]dx_\mathrm{p}. \label{eq:ENvL}
\end{multline}
\hrulefill
\end{figure*}

\section{PA Position Selection via REMs} 
\label{sec:REM}
As discussed in Sec. \ref{sec:stat_char}, the slowly varying and stable nature of long-term power makes a \gls{REM} an effective strategy to identify suitable \gls{PA} positions based on long-term power. Importantly, their validity does not depend on channel stationarity over \gls{SSF} timescales, and they remain valid provided the large-scale propagation environment does not change significantly over the timescale of REM construction and use.

An \gls{REM} can be constructed in a range of ways, but here we focus on a low-complexity method. We consider a random-look approach that serves as a lower bound, where $\beta(x_\mathrm{p})$ is estimated opportunistically at locations where a \gls{UE} happens to be, and the resulting samples are spatially averaged to gradually form a more accurate grid over time. We then consider a perfect grid-based approach, where $\beta(x_\mathrm{p})$ is estimated at specific predetermined locations, forming a grid of \gls{UE} positions. This approach represents an idealized upper bound scenario in which measurements of the environment are taken at exactly equal intervals. We consider both nearest-neighbor and interpolation-based \gls{PA} position selection methods, and consider localization error. 

Previously, $h_\mathrm{p}(x_\mathrm{p})$ denoted the channel between the \gls{UE} at $\boldsymbol{\psi}_\mathrm{u}$ and the \gls{PA}. Here, with multiple UE locations involved, we write $h_\mathrm{p}(x_\mathrm{p}, \boldsymbol{\psi}_\mathrm{u})$ to make the dependence explicit.



\subsection{Random Look-based Construction}
In this scenario, the \gls{REM} is constructed from a large sample of random looks. Each look is an estimate of the channel power between the \gls{PA} at a random position $x_\mathrm{p}\in[0,L]$ and a \gls{UE} at some location $(x_\mathrm{u},y_\mathrm{u})\in[0,D_x]\times[0,D_y]$, where $D_x$ and $D_y$ are the dimensions in meters of the environment considered. This approach is simple and  practical, as these measurements can be taken at any time as a user moves naturally through the environment, resulting in minimal additional estimation overhead. As more measurements are accumulated, the \gls{REM} becomes progressively more accurate. In principle, given an infinite number of samples, a continuous surface of channel knowledge could be formed, and performance would approach the true instantaneous optimum, rather than a spatially-averaged result. In practice, samples are aggregated within discretized spatial averaging regions so that as more samples are collected within a given region, its running average converges towards the true LSF value for that region. Therefore, performance approaches an ideal grid-based REM result, which represents the best achievable outcome. This upper bound is discussed in Sec. \ref{sec:gbconst}.

For each random observation, the total channel power is measured and averaged over $\tau$ symbols, during which the channel is assumed to remain constant. Using least squares, the estimated channel power between the \gls{UE} and the \gls{PA} is  
\begin{align}
    |\hat{h}_\mathrm{p}(\mathbf{z})|^2=&\tfrac{1}{\tau E_s}\sum\nolimits_{k=1}^\tau|r_k(\mathbf{z})|^2-\tau\sigma^2, \notag \\
    =&\tfrac{1}{\tau E_s}\Big[\sum\nolimits_{k=1}^\tau|{h}_\mathrm{p}(\mathbf{z})|^2|s_k|^2+\sum\nolimits_{k=1}^\tau|n_k|^2 \notag \\ &\qquad\quad+2\sum\nolimits_{k=1}^\tau\Re\{{h}_\mathrm{p}(\mathbf{z})s_kn_k^*\}-\tau\sigma^2\Big],
    \label{eq:estchanpower}
\end{align}
where $\mathbf{z}=(\boldsymbol{\psi}_{\mathrm{u}},x_\mathrm{p})$, $r_k(\mathbf{z})=h_\mathrm{p}(\mathbf{z})s_k+n_k$ is the received signal, $s_k$ is the $k$-th training symbol with energy $E_s$, and $n_k$ is the noise with variance $\sigma^2$. The cross product terms in \eqref{eq:estchanpower} have a mean of zero, and the impacts of noise variance and signal strength are compensated for, giving an unbiased estimate of channel power. However, these samples are at a single point of space and time. To construct a \gls{REM} that captures long-term channel behavior, a spatial averaging procedure is employed.
Each \gls{REM} data point corresponds to a grid point $\mathbf{z}_s=(\boldsymbol{\psi}_{\mathrm{u},s}, x_{\mathrm{p},s})$. At each such grid point, the estimated long-term SNR, $\hat{\mathrm{SNR}}_\mathrm{LT}(\mathbf{z}_s)$, is obtained by spatially averaging all received SNR samples collected at \gls{PA} positions within a distance $\delta_\mathrm{p}$ of $x_{\mathrm{p},s}$ from \gls{UE} positions within a radius $\delta_\mathrm{e}$ of $\boldsymbol{\psi}_{\mathrm{u},s}$. The full random look-based \gls{REM} construction process is detailed in Algorithms \ref{alg:RREM} and \ref{alg:SpAv}.

\begin{algorithm}[ht]
\caption{Random look-based \gls{REM} construction}
\label{alg:RREM}
Set number of observations, $N$, and samples, $\tau$, spatial averaging radii, 
$\delta_\mathrm{p}$ and $\delta_\mathrm{e}$, grid spacings $\Delta_\mathrm{p}$, $\Delta_x$ and $\Delta_y$. Initialize $n \gets 1$\; 
\While{$n \leq N$}{
    Request a sample from a random \gls{UE}\;
    Select a random \gls{PA} position $x_{\mathrm{p},n}$\;
    Receive $\tau$ training data symbols\;
    Receive location data $\boldsymbol{\psi}_{\mathrm{u},n}=(x_{\mathrm{u},n}, y_{\mathrm{u},n})$\;
    Compute channel power $|\hat{h}_\mathrm{p}(\mathbf{z}_n)|^2$ via \eqref{eq:estchanpower}\;
    $n \gets n+1$\;
    }
Construct REM with $\hat{\mathrm{SNR}}_\mathrm{LT}(\mathbf{z}_s)$ values from Alg. \ref{alg:SpAv} at spatial averaging grid points $\mathbf{z}_s=(x_{\mathrm{u},s}{=}1\!:\!\Delta_x\!:\!D_x$, $\!y_{\mathrm{u},s}{=}1\!:\!\Delta_y\!:\!D_y$, $x_{\mathrm{p},s}{=}1\!:\!\Delta_\mathrm{p}\!:\!L)$.
\end{algorithm}
\begin{algorithm}[ht]
\caption{Spatial averaging to remove fast fading}
\label{alg:SpAv}
    \For{$x_{\mathrm{u},s}{=}1\!:\!\Delta_x\!:\!D_x$, $\!y_{\mathrm{u},s}{=}1\!:\!\Delta_y\!:\!D_y$, $\!x_{\mathrm{p},s}{=}1\!:\!\Delta_\mathrm{p}\!:\!L\!$}{
$\mathbf{z}_s=(x_{\mathrm{u},s}, y_{\mathrm{u},s}, x_{\mathrm{p},s})=(\boldsymbol{\psi}_{\mathrm{u},s}, x_{\mathrm{p},s})$\;
    $\mathcal{S}_s(\mathbf{z}_{s})\!=\!\! \{n\!:\! |x_{\mathrm{p},s}\!-x_{\mathrm{p},n\!}|\! \le\! \delta_\mathrm{p},\!\|\boldsymbol{\psi}_{\mathrm{\!u},s}\!\!-\boldsymbol{\psi}_{{\mathrm{\!u}},n\!}\|_2 \!\le\! \delta_\mathrm{e} \}$\;
        $\hat{\mathrm{SNR}}_\mathrm{LT}(\mathbf{z}_s)
        = \frac{1}{|\mathcal{S}_{s}|}\sum_{n\in\mathcal{S}_{s}}|\hat{h}_\mathrm{p}(\mathbf{z}_n)|^2$\;
    }
\end{algorithm}
This process can occur continuously over time, allowing the \gls{REM} to evolve as new measurements are collected and spatial averages updated. As a result, the effective sampling density around each grid point increases, yielding progressively more accurate power estimates.

The overheads associated with the random look-based \gls{REM} are low, and we summarize them in the list below:
\begin{itemize}
    \item \textbf{Measurement overhead:} Data collection is opportunistic, relying on channel measurements taken during normal system operation rather than a dedicated measurement campaign. Each observation consists of a single channel measurement at the reporting \gls{UE}'s location, with no additional measurement procedure required.
    \item \textbf{Computational overhead:} Offline processing per observation involves computing the channel gain, removing fast-fading fluctuations, and updating the averaged value for the corresponding region. This is a low-complexity, $O(1)$ operation per sample, requiring only a weighted running average rather than reprocessing of previously collected data. 
    \item \textbf{Storage overhead:} Storage scales with the number of discretized regions in the \gls{REM}, not with the number of samples collected. Once a sample has been incorporated into a region's running average, it can be discarded, so the storage burden remains constant regardless of the volume of data collected.
    \item \textbf{Update overhead:} Updating a region's value on receipt of a new sample requires only the previous averaged value, the previous sample count, and the new observation, giving an update cost that is independent of the amount of historical data.
\end{itemize}

\glspl{UE} will not be located exactly on a grid point. For the random look-based \gls{REM}, we employ the simplest approach to selecting a \gls{PA} position, which is to choose the grid point nearest to the \gls{UE}. We refer to this approach as the \gls{NN} method, for which the selected PA position is 
\begin{equation}
x_\mathrm{p,NN}^\star = \arg\min_{x_{\mathrm{p},s}} \| \boldsymbol{\psi}_{\mathrm{u},s} - \boldsymbol{\psi}_\mathrm{u} \|_2 . \label{eq:NN}
\end{equation}

\subsection{Perfect Grid-based Construction}
\label{sec:gbconst}
Here, we consider the construction and use of an ideal grid-based \gls{REM}, which produces a balanced map across all regions of the environment. The environment is partitioned into a two-dimensional grid of $N_x \times N_y$ points, with spacing $\Delta_x = D_x/(N_x-1)$ and $\Delta_y = D_y/(N_y-1)$ to cover the entire area. The \gls{PA} \gls{WG} is also discretized into $N_{\mathrm{p}}$ sample points with spacing $\Delta_\mathrm{p} = L/(N_\mathrm{p}-1)$. As it is unlikely that UEs will be located precisely at grid points in practice, the grid-based REM does not correspond to a realistic deployment; it instead provides a bound against which the random look-based REM can be compared, representing the case where samples perfectly cover the environment. As this is intended purely as an idealized bound, channel estimation overheads were not considered for this approach.

The channel power is estimated at $\mathbf{z}_s\!=\!(x_{\mathrm{u},s}, y_{\mathrm{u},s}, x_{\mathrm{p},s})$ for all $x_{\mathrm{u},s}, y_{\mathrm{u},s}, x_{\mathrm{p},s}$ as in \eqref{eq:estchanpower}, resulting in a 3-dimensional structure of $N_x \times N_y \times N_{\mathrm{p}}$ measurements. The \gls{PA} sample index that yields the maximum channel power at each grid point is then extracted, producing an $N_x \times N_y$ map that associates each physical location with the best-performing \gls{PA} position. This process is shown in Algorithm \ref{alg:GREM}. 

\begin{algorithm}[ht]
Set numbers of grid points $N_\mathrm{p}$, $N_x$ and $N_y$, and compute grid spacings $\Delta_\mathrm{p}$, $\Delta_x$, and $\Delta_y$\;
\For{$x_{\mathrm{u},s}{=}1\!:\!\Delta_x\!\!:\!D_x$, $\!y_{\mathrm{u},s}{=}1\!:\!\Delta_y\!\!:\!D_y$, $\!x_{\mathrm{p},s}{=}1\!:\!\Delta_\mathrm{p}\!\!:\!L$}{
    $\mathbf{z}_s=(x_{\mathrm{u},s}, y_{\mathrm{u},s}, x_{\mathrm{p},s})=(\boldsymbol{\psi}_{\mathrm{u},s}, x_{\mathrm{p},s})$\;
        Receive $\tau$ training data symbols for $\mathbf{z}_s$\;
        Compute channel power $|\hat{h}_\mathrm{p}(\mathbf{z}_s)|^2$ via \eqref{eq:estchanpower}\;
}
Compute $\hat{\mathrm{SNR}}_\mathrm{LT}(\mathbf{z}_s)$ values from Alg. \ref{alg:SpAv}\;
Construct \gls{REM} from $\mathbf{z}_s^\star=(\boldsymbol{\psi}_{\mathrm{u},s},x^\star_{\mathrm{p},s})$ values, where $x^\star_{\mathrm{p},s}=\arg\max_{x_{\mathrm{p},s}}\hat{\mathrm{
SNR}}_\mathrm{LT}(\mathbf{z}_s)$.
\caption{Grid-based \gls{REM} construction}
\label{alg:GREM}
\end{algorithm}

\begin{figure}[ht]
    \centering
    \includegraphics[width=0.9\linewidth]{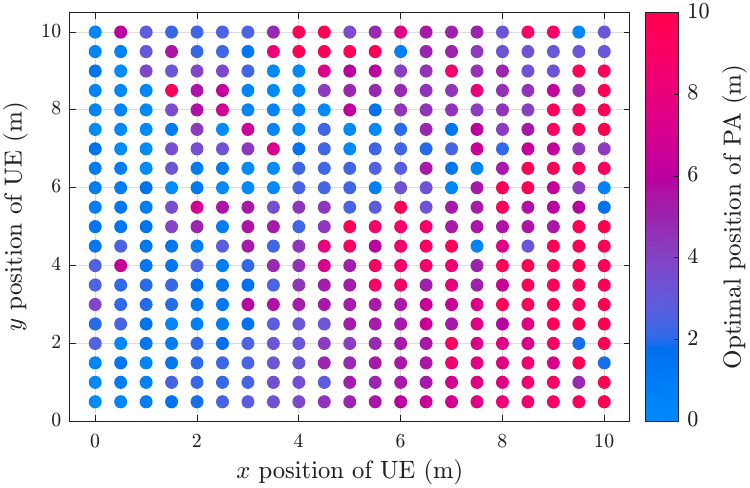}
    \caption{A simulated grid-based \gls{REM} for points spaced 0.5 m apart in a $10\times 10$ m$^2$ environment, where the point color indicates the best position on a 10 m long \gls{PA} located along the $x$-axis.}
    \label{fig:gridREM}
\end{figure}

An example of a simulated grid-based \gls{REM} is shown in Fig. \ref{fig:gridREM} for a system where $f_c=28$ GHz, $\alpha=0.05$, $\gamma=2.2$, $\sigma_\mathrm{SF}=8$~dB, $K=1$, $\kappa=0.1$ dB/m and $n_\mathrm{eff}=1.4$ \cite{ding_flexible_2025}.

A sensitivity analysis of the impact of localization errors during \gls{REM} construction can be performed by incorporating an error term into the chosen \gls{PA} positions. When estimating the channel power at a \gls{WG} location $x_{\mathrm{p},s}$, the actual channel is assumed to correspond to a nearby position $x_{\mathrm{p},s} + e$, where $e \sim \mathcal{N}(0,\sigma_\mathrm{e}^2)$ represents position uncertainty. As this error model is generic, it can represent other practical imperfections as well. For example, outdated \gls{UE} location or channel information is equivalent to information collected at an incorrect position, as the system has since moved on. Therefore, this behavior can be captured by the general error term $e$. Note that $\sigma_\mathrm{e}$ is defined in index units, so the corresponding error has a standard deviation of $\sigma_\mathrm{e} \Delta_\mathrm{p}$. Results both with and without the error term $e$ will be shown in Sec \ref{sec:numresREM}.

We consider two \gls{PA} position selection methods for the perfect grid-based \gls{REM}: the \gls{NN} approach in \eqref{eq:NN}, and an interpolation-based approach which averages the four nearest grid points to obtain a \gls{PA} position.\footnote{We consider both PA position selection methods only in conjunction with the perfect grid-based \gls{REM}, as the ideal REM allows the impact of different selection strategies to be isolated from the accuracy of the REM itself.} If the \gls{LSF} is perfectly Gaussian, the optimal \gls{PA} position for a \gls{UE} located at $\boldsymbol{\psi}_\mathrm{u}$ is
\begin{equation}
    x^\star_\mathrm{p,GI}{=}\arg\max_{x_\mathrm{p}}\mathbb{E}\!\big[{\mathrm{SNR}}_\mathrm{LT,dB}(\mathbf{z})|\{\hat{\mathrm{SNR}}_\mathrm{LT,dB}(\mathbf{z}^\star_{i})\}_{i=1}^4\!\big]\!,
\end{equation}
where $\{\mathbf{z}^\star_{i}\}^4_{i=1}$ denotes the \gls{REM} data at the four grid points surrounding the \gls{UE} and $\mathrm{SNR}_\mathrm{LT,dB}(\mathbf{z})=10\log_{10}(\mathrm{SNR}_\mathrm{LT}(\mathbf{z}))$ and $\hat{\mathrm{SNR}}_\mathrm{LT,dB}(\mathbf{z})=10\log_{10}(\hat{\mathrm{SNR}}_\mathrm{LT}(\mathbf{z}))$ are the exact and estimated long-term \glspl{SNR} in dB, which are distance-dependent Gaussians. However, this approach requires knowledge of all covariances, and the \gls{LSF} will likely not be perfectly Gaussian. We instead investigate a more pragmatic \gls{BI}-based method. Let $l_\mathrm{f}{=}\lfloor\frac{x_\mathrm{u}}{\Delta_x}\rfloor$ and $m_\mathrm{f}{=}\lfloor\frac{y_\mathrm{u}}{\Delta_y}\rfloor$, such that $w_x{=}\frac{x_\mathrm{u}-x_{\mathrm{u},l_\mathrm{f}}}{x_{\mathrm{u},l_\mathrm{f}+1}-x_{\mathrm{u},l_\mathrm{f}}}$ and $w_y{=}\frac{x_\mathrm{u}-x_{\mathrm{u},m_\mathrm{f}}}{x_{\mathrm{u},m_\mathrm{f}+1}-x_{\mathrm{u},m_\mathrm{f}}}$. Letting $x_{\mathrm{p},l_\mathrm{f},m_\mathrm{f}}$ denote the \gls{PA} position saved for \gls{REM} grid point $(l_\mathrm{f}, m_\mathrm{f})$, the \gls{PA} position can be set as 
\begin{multline}
    x_\mathrm{p,BI}^\star\!=\!(1\!-\!w_x)(1\!-\!w_y)x_{\mathrm{p},l_\mathrm{f},m_\mathrm{f}}\!+\!w_x(1\!-\!w_y)x_{\mathrm{p},l_\mathrm{f}+1,m_\mathrm{f}} \\ +\!(1\!-\!w_x)w_yx_{\mathrm{p},l_\mathrm{f},m_\mathrm{f}+1}\!+\!w_xw_yx_{\mathrm{p},l_\mathrm{f}+1,m_\mathrm{f}+1}.
\end{multline}
The performance of the \gls{NN} and \gls{BI} position selection methods for the perfect grid-based \gls{REM} will be compared in Sec. \ref{sec:numresREM}.

\section{Numerical Results}
\label{sec:numres}
Numerical results were generated to verify the analysis in Secs. \ref{sec:expcorr_analysis} and \ref{sec:smoothcorr_analysis} and the performance of the \glspl{REM} proposed in Sec. \ref{sec:REM}. The attenuation factor along the dielectric is $\kappa=0.1$ dB/m and the effective refractive index of the dielectric is $\eta_\mathrm{eff}=1.4$ \cite{ding_flexible_2025}. We consider a path loss exponent of $\gamma=2.2$ and a shadow fading standard deviation of $\sigma_\mathrm{SF}=8$ dB \cite{goldsmith_wireless_2005}. Fast fading samples are taken every $10$ cm. Unless otherwise specified, it is assumed that the Ricean $K$-factor is $K=1$, the carrier frequency is $f_\mathrm{c}=28$ GHz and the spatial decorrelation distance is $d_\mathrm{LS}=20$ m \cite{molisch_wireless_2011}, corresponding to $\alpha=0.05$. All other parameters vary between the different scenarios considered so are specified for each figure.

\subsection{Exponentially Correlated LSF}
In this section, we investigate the performance of the mean and distribution of $\mathrm{SNR_{TS}}$ derived in Sec.~\ref{sec:expcorr_analysis}. We assume that $\boldsymbol{\psi}_\mathrm{u}=(1/2, 5)$. This does not compromise the generality of the conclusions. As the \gls{UE} moves towards the end of the \gls{WG}, the resulting \gls{PDF} of the \gls{SNR} shifts slightly to the left while retaining the same overall shape. This is because fewer candidate \gls{PA} positions lie nearby, increasing the average distance to the selected position and thus the path loss, slightly reducing the optimal \gls{SNR}. This effect grows with \gls{WG} length but remains small even for long \glspl{WG}.

\subsubsection{SNR distribution}

\begin{figure}[t]
    \centering
    \includegraphics[width=0.9\linewidth]{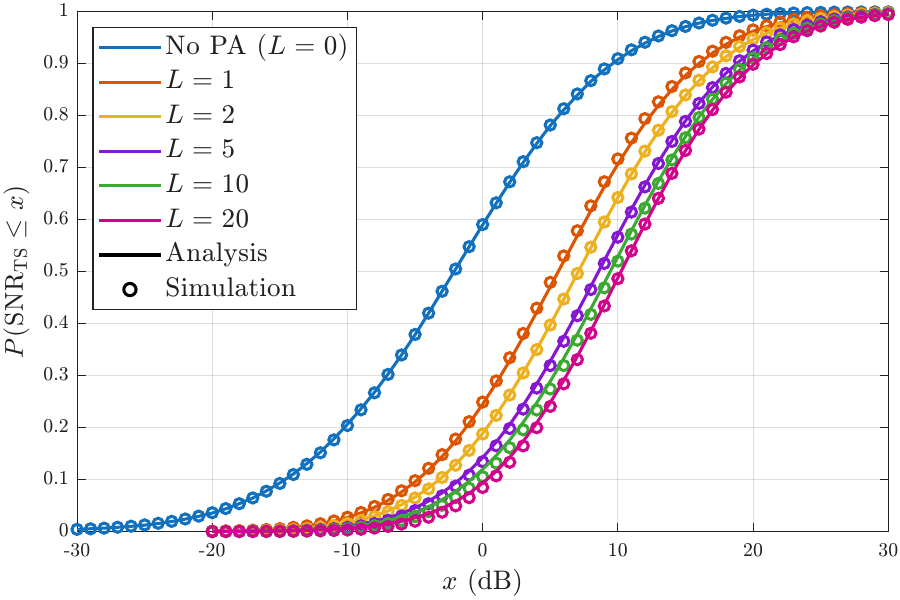}
    \caption{The distribution of $\mathrm{SNR_{TS}}$ as stated in \eqref{eq:SNRmaxdist} for a range of \gls{PA} lengths.}
    \label{fig:expcorr}
\end{figure}

Figure \ref{fig:expcorr} investigates the impact of \gls{WG} length $L$ on the distribution of $\mathrm{SNR_{TS}}$ derived in Sec. \ref{sec:expcorr_analysis_mean}. It also compares the performance of a \gls{PA}-assisted system with a traditional system, where transmission occurs from an antenna fixed at $x_\mathrm{p}=0$ to the \gls{UE}. We set $\frac{E_s}{\sigma^2}$ such that $P(\mathrm{SNR}_\mathrm{TS}>5\,\, \mathrm{dB})=0.5$ for $K=1$ and $f_\mathrm{c}=28$~GHz.

The derived \gls{SNR} distribution exhibits high accuracy over the entire range, underscoring its usefulness for system analysis. As the \gls{PA} is introduced and the \gls{WG} length increases, the curves shift to the right, indicating that \glspl{PA} with longer \glspl{WG} can yield improved performance in scattered channels. Increasing the \gls{WG} length creates more spatial locations over which both the slowly varying large-scale power and the small-scale scattered components can be effectively exploited for diversity. The performance gains obtained by increasing $L$ diminish as $L$ becomes larger. This behavior arises because substantial spatial diversity is already present, rendering the additional diversity provided by further increases in length less impactful. For example, adding a $L=1$ m \gls{PA} to a traditional system leads to a $7.7$ dB improvement  in the median \gls{SNR}, increasing $L$ from $1$ m to $2$ m yields a $1.5$ dB improvement, while increasing $L$ from $10$ m to $20$ m results in a gain of $0.8$ dB. The performance gains from extending the \gls{WG} are more substantial when the scattered component is stronger, as the additional diversity can enhance a larger proportion of the overall channel. This highlights the importance of accounting for the Ricean \(K\)-factor when selecting the \gls{WG} length in a \gls{PASS}.

\subsubsection{Mean SNR}
\begin{figure}[ht]
    \centering
    \includegraphics[width=0.9\linewidth]{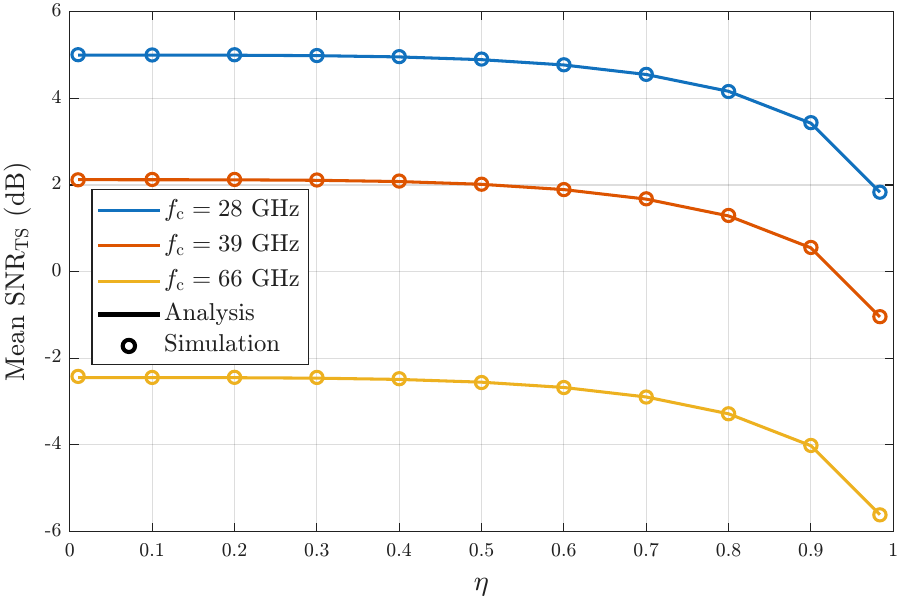}
    \caption{The mean $\mathrm{SNR_{TS}}$ for a range of $K$-factors and carrier frequencies.}
    \label{fig:meanSNR}
\end{figure}

Figure \ref{fig:meanSNR} compares the mean $\mathrm{SNR_{TS}}$ result in \eqref{eq:ESNRmax} with simulations for a range of Ricean $K$-factors and carrier frequencies. Note that $\eta=\sqrt{\frac{K}{K+1}}$ rather than $K$ is plotted on the $x$-axis to clearly show the proportion of the channel that is \gls{LoS}. We assume that $L=1$ and set $\frac{E_s}{\sigma^2}$ such that $\mathbb{E}[\mathrm{SNR}_\mathrm{TS}]=5$~dB for $K=1$ and $f_\mathrm{c}=28$~GHz.

The derived expression is virtually indistinguishable from the simulated results, highlighting its accuracy as an analytical tool. When $\eta$ is small (corresponding to a low Ricean $K$-factor), the scattered component is sufficiently strong to induce pronounced fast-fading fluctuations. Since the maximum \gls{SNR} is considered and the best channel realization is selected, these fluctuations contribute positively, leading to an increase in the mean of $\mathrm{SNR_{TS}}$. As the Ricean $K$-factor increases and the link becomes predominantly \gls{LoS}, the channel fluctuations are suppressed, and the mean of $\mathrm{SNR_{TS}}$ consequently decreases.  However, this decrease is accompanied by a lower variance, yielding more stable \gls{SNR} values. Also, $\mathbb{E}[\mathrm{SNR_{TS}}]$ decreases with increasing $f_\mathrm{c}$ as free-space path loss scales with $f_\mathrm{c}^2$.

System design priorities depend on the desired performance objectives. To maximize the peak \gls{SNR}, lower carrier frequencies and moderate Ricean $K$-factors are advantageous (provided that the long-term \gls{SNR} is not severely degraded), as they enhance the contribution of scattered components and the resulting fast-fading variability. For reliability and stable \gls{SNR}, higher $K$-factors are preferable, since they suppress fading fluctuations. Conversely, higher carrier frequencies provide larger available bandwidth, at the expense of a lower mean \gls{SNR} caused by increased propagation losses.

\subsection{Smoothly Correlated Large-Scale Fading}
In this section, we investigate the performance of the instantaneous \gls{SNR} distribution derived in Sec.~\ref{sec:smoothcorr_analysis} under Gaussian-correlated shadow fading. Specifically, we assume that $\mathbb{E}[X(x_\mathrm{p})X(x_\mathrm{p}+\tau)]=\mathrm{exp}({-(\alpha\tau)^2})$.

\subsubsection{SNR Distribution}
\begin{figure}[ht]
    \centering
    \includegraphics[width=0.9\linewidth]{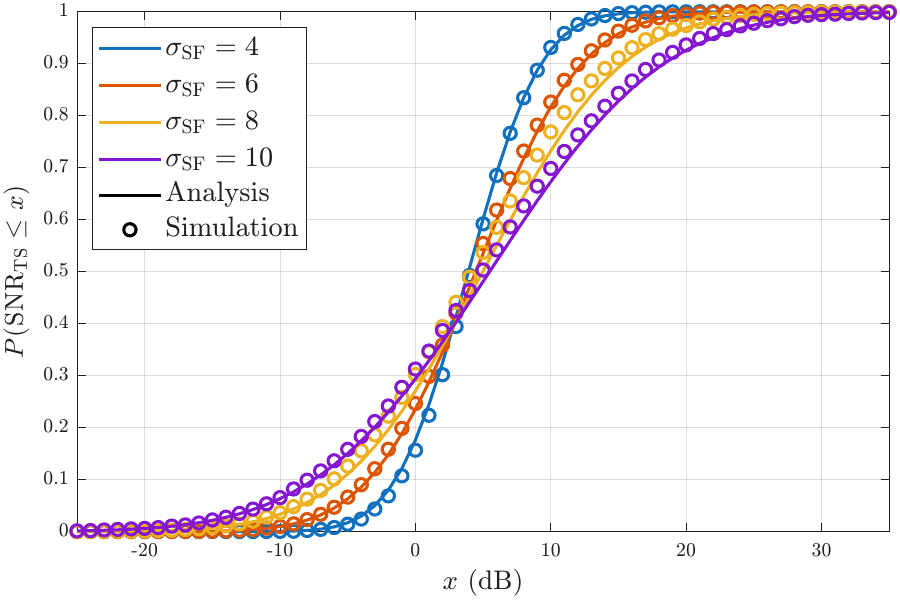}
    \caption{The \gls{SNR} distribution in \eqref{eq:SNRmaxdist} for a range of shadow fading standard deviations.}
    \label{fig:smoothcorrOP}
\end{figure}

Figure \ref{fig:smoothcorrOP} investigates the \gls{SNR} distribution for a range of different shadow fading standard deviations. Here, we assume that $L=5$, $\boldsymbol{\psi}_\mathrm{u}=(L/2, 5)$, and $\frac{E_s}{\sigma^2}$ is set such that $P(\mathrm{SNR}_\mathrm{TS}>5\,\, \mathrm{dB})=0.5$ when $\sigma_\mathrm{SF}=8$ for $K=1$ and $f_\mathrm{c}=28$~GHz.

Figure \ref{fig:smoothcorrOP} shows that the upper tail of the distribution increases with \(\sigma_\mathrm{SF}\), indicating that higher shadow fading variance can be beneficial for system performance. However, unlike fast fading variance, increasing \(\sigma_\mathrm{SF}\) also increases the outage probability due to the correlated nature of the shadow fading. For independent realizations, such as sufficiently spaced fast fading samples, higher variance shifts the entire distribution to the right, similar to the effect observed in Fig.~\ref{fig:expcorr}, where increased \gls{WG} length produced more samples and a higher probability of achieving a very high \gls{SNR}. In contrast, when the samples are correlated, a large variance can cause all locations to simultaneously exhibit low or high \gls{SNR} in some realizations, exaggerating both extremes. Consequently, high $\sigma_\mathrm{SF}$ can improve the upper tail but worsen outage probability. With low spatial correlation (i.e., shorter decorrelation distance), a high $\sigma_\mathrm{SF}$ can positively affect the entire $\mathrm{SNR_{TS}}$ distribution.

\subsection{REM Performance}
\label{sec:numresREM}
In this section, we investigate the performance of the \gls{REM} construction and position selection methods proposed in Sec. \ref{sec:REM}. The exponential correlation model in \eqref{eq:exp_corr} is used for all \gls{REM} results. Here, $L=20$ m, the \gls{PA} is discretized into 100 positions that can be selected, and $\boldsymbol{\psi}_\mathrm{u}=(x_\mathrm{u}, y_\mathrm{u})$ is drawn randomly, where $x_\mathrm{u}\in[0,L]$ and $y_\mathrm{u}\in[0,L]$. Note that $\frac{E_s}{\sigma^2}$ is set so that the median value of $\mathrm{SNR_{LT}}$ is $5$ dB for an optimally located \gls{PA} when $K=1$ and $f_\mathrm{c}=28$ GHz. In these results, REM-GBNN, REM-GBNNE and REM-GBI refer to the grid-based REM with the \gls{PA} position selected by the \gls{NN} method, with and without localization error, and the interpolation method, respectively. In all cases, there are 10 grid points in both the $x$ and $y$ directions equally spaced between 0 and $L$. REM-RL ($x$) denotes the random look-based construction method where $x$ is the number of looks considered. Random is a randomly selected \gls{PA} point that acts as a lower bound, while optimal refers to a finer version of the REM-GBNN with 8 additional grid points located between adjacent REM points, giving near optimal performance.

\subsubsection{Comparison of Construction Methods}
\begin{figure}[ht]
    \centering
    \includegraphics[width=0.9\linewidth]{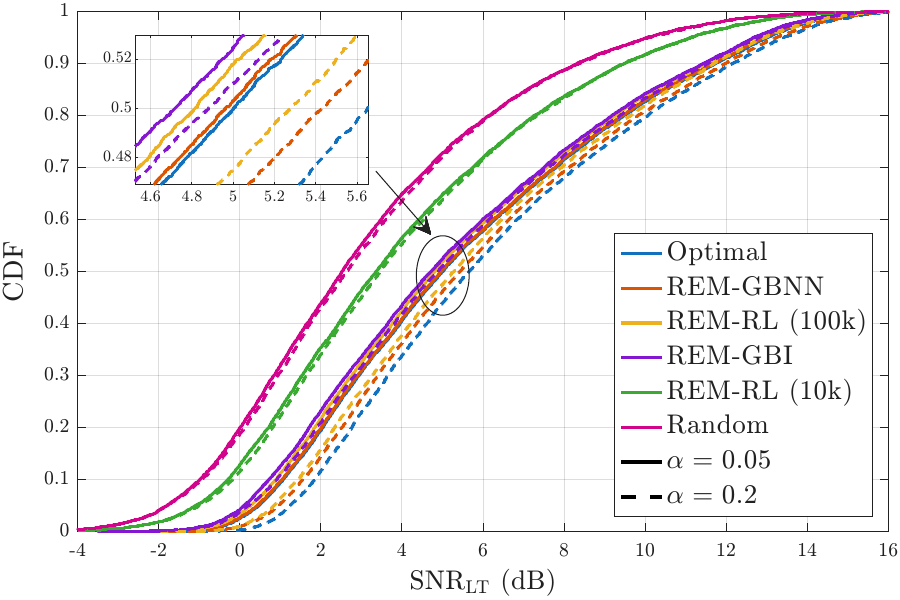}
    \caption{The CDF of the long-term SNR for a range of REM-based PA position selection methods compared to random and optimal baselines.}
    \label{fig:REM_construc}
\end{figure}

Figure \ref{fig:REM_construc} compares the performance of the \gls{REM} methods proposed in Sec.~\ref{sec:REM} against a random lower bound and a near-optimal upper bound across $10,000$ different \gls{UE} locations for decorrelation distances of $d_\mathrm{LS}=5$ m and $20$ m.

The grid-based \gls{NN} approach achieves the best performance for both $\alpha$ values, followed by the random-look approach with 100,000 samples. Among the grid-based methods, \gls{NN} position selection outperforms interpolation-based selection due to the effects of fading. The four points used for interpolation may differ significantly because of fading, causing the interpolated value to either deviate from all of them or fall within a deep fade. This is a useful and generalizable insight: for REM value selection, using the nearest value can be more effective than interpolating across many nearby points. Importantly, this finding is not specific to the grid-based REM construction method considered here. The ideal construction method was used for these results solely to isolate the effect of REM value selection strategy from construction artifacts. Additionally, the effect of the number of samples in the random-look approach is evident:  10,000 samples yields performance close to that of random position selection, whereas 100,000 samples approaches the optimal result. If the number of samples continued to increase, the performance would eventually converge to the optimal. 

\begin{figure}[t]
  \centering
  \subfloat[Near-optimal $\mathrm{SNR_{LT}}$ ($\alpha{=}0.05$).]{%
    \includegraphics[width=0.475\linewidth]{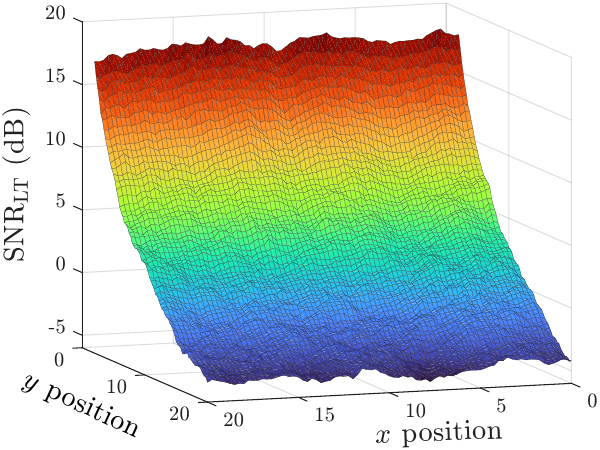}
    \label{fig:sub1}
  }
  \hfill
  \subfloat[Near-optimal $\mathrm{SNR_{LT}}$ ($\alpha{=}0.2$).]{%
    \includegraphics[width=0.475\linewidth]{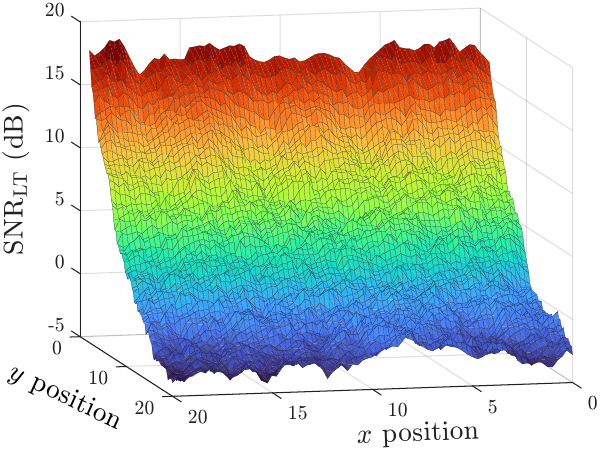}
    \label{fig:sub2}
  }
  \hfill
  \subfloat[$\mathrm{SNR_{LT}}$ at REM points ($\alpha{=}0.05$).]{%
    \includegraphics[width=0.475\linewidth]{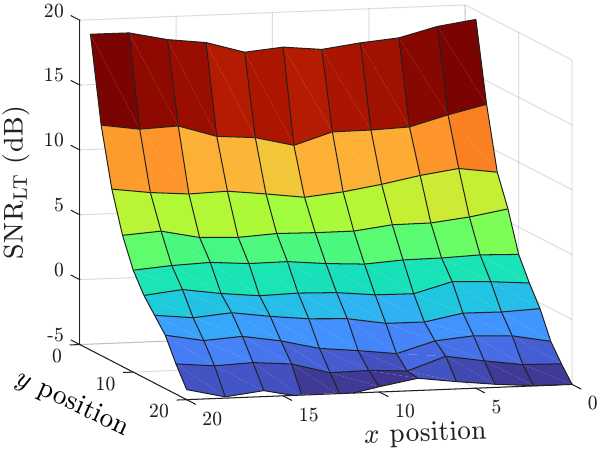}
    \label{fig:sub3}
  }
    \hfill
  \subfloat[$\mathrm{SNR_{LT}}$ at REM points ($\alpha{=}0.2$).]{%
    \includegraphics[width=0.475\linewidth]{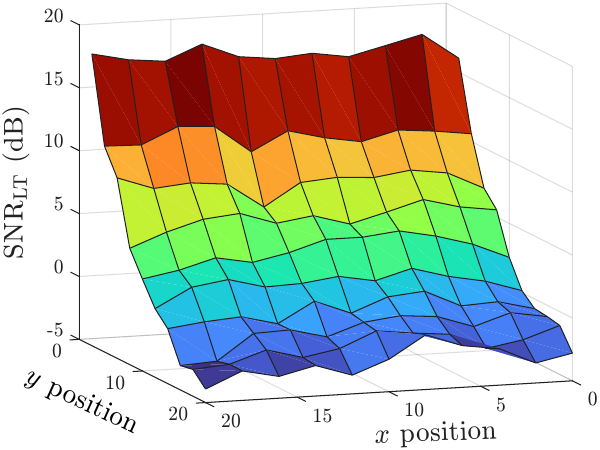}
    \label{fig:sub4}
  }
  \caption{Example of the optimal long-term SNR and the long-term SNR at REM grid points for $\alpha=0.05$ and $\alpha=0.2$.}
  \label{fig:REMsurface}
\end{figure}

The long-term \gls{SNR} performance improves as the decorrelation distance decreases. For all proposed methods, the \gls{REM} is constructed using the maximum value either at a grid point or within a small radius, and the increased randomness associated with lower spatial correlation is beneficial for maximizing the \gls{SNR}. However, the proposed methods achieve results closer to the optimal when the correlation is high. This is because the \gls{SNR} varies more slowly in space under high correlation, causing measurements taken at nearby positions to more accurately reflect the true maximum. This effect is shown in Fig. \ref{fig:REMsurface}, where the maximum $\mathrm{SNR_{LT}}$ value at each \gls{REM} grid point and near-optimal fine grid point for $\alpha=0.2$ ($d_\mathrm{LS}=5$ m) and $\alpha=0.05$ ($d_\mathrm{LS}=20$ m) is illustrated.

The surface in Fig.~\ref{fig:sub2}, where \(d_\mathrm{LS}=5\)~m, varies much more rapidly with distance than the surface in Fig.~\ref{fig:sub1}, where \(d_\mathrm{LS}=20\)~m. The increased spatial variability causes the corresponding \gls{REM} to miss important changes between its comparatively coarse grid points, thereby reducing its effectiveness. In contrast, the \gls{REM} shown in Fig.~\ref{fig:sub3} closely matches the smooth optimal surface, indicating that the nearest-neighbor approach is more likely to identify the optimal \gls{PA} position. Additionally, the values used for interpolation are more similar, leading to more accurate position estimates.  

The performance of the random-look approach remains relatively consistent across different $d_\mathrm{LS}$ values, as it relies less on local spatial similarity and more on the number of samples. Thus, grid-based \gls{REM} constructions are better suited to scenarios with a large \(d_\mathrm{LS}\), while random-look-based constructions are a more effective choice when \(d_\mathrm{LS}\) is small.

\subsubsection{Impact of Positioning Error}

\begin{figure}[t]
    \centering
    \includegraphics[width=0.9\linewidth]{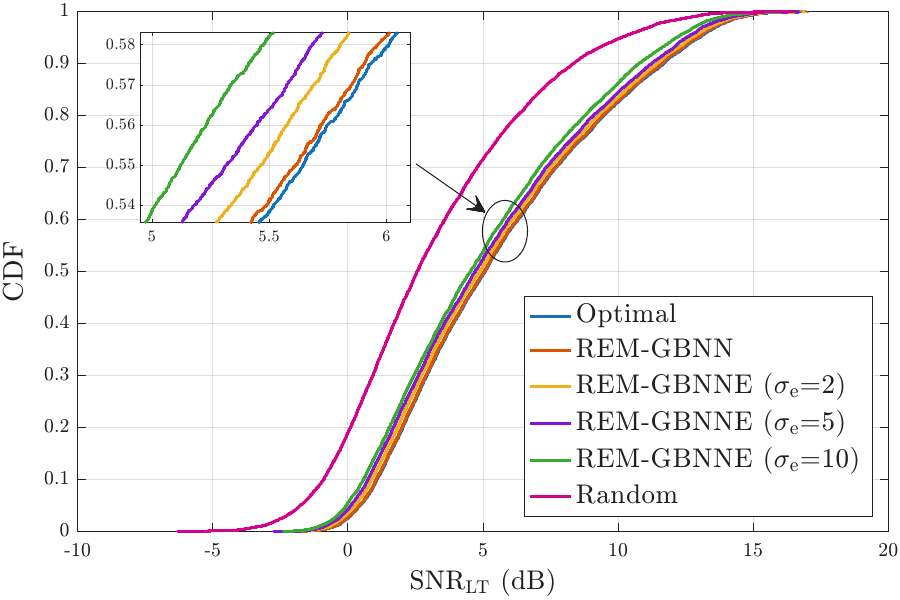}
    \caption{The CDF of the long-term SNR for the grid-based REM PA position selection method with localization error variances of $\sigma_\mathrm{e}=\{2,5,10\}$ compared against random, optimal, and error-free grid-based REM baselines. }
    \label{fig:REM_chanest}
\end{figure}

Figure \ref{fig:REM_chanest} shows the impact of localization error on the grid-based \gls{REM} using \gls{NN} \gls{PA} position selection. Multiple error standard deviation values, $\sigma_\mathrm{e}$, are considered. 

Figure \ref{fig:REM_chanest} illustrates that the grid-based \gls{REM} method is highly robust to errors. Even for $\sigma_\mathrm{e}=10$, the median long-term \gls{SNR} decreases by only $0.58$ dB relative to the optimal, with negligible changes in the \gls{CDF} tails. This resilience stems from the channel’s spatial correlation. When the selected location deviates from the intended one due to errors from imperfect localization, outdated channel state information etc., nearby locations exhibit similar channel conditions, reducing the impact on performance. These results suggest that, in practical REM deployments, moderate errors would have little impact on long-term \gls{SNR}, highlighting that \glspl{REM} are particularly well-suited for real-world implementation.

\subsubsection{Comparison with Benchmarks}
\begin{figure}[t]
    \centering
    \includegraphics[width=0.9\linewidth]{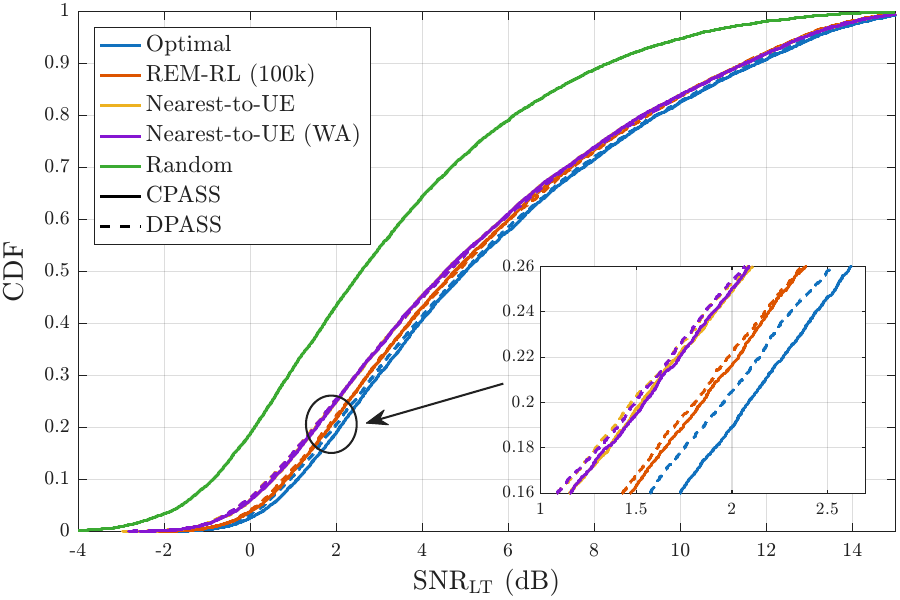}
    \caption{The \gls{CDF} of the long-term \gls{SNR} for the random look-based \gls{REM} compared to optimal, nearest-to-\gls{UE} with and without accounting for \gls{WG} attenuation (WA) and random positioning benchmarks, for both a continuous \gls{PASS} (CPASS) and a discrete \gls{PASS} (DPASS).}
    \label{fig:REM_bench}
\end{figure}

Figure \ref{fig:REM_bench} compares the performance of the random look-based \gls{REM} with two closest-to-\gls{UE} benchmarks. As the majority of \gls{PASS} literature considers purely \gls{LoS} channels and therefore neglects shadow fading, one of the most widely adopted PA positioning strategies sets $x_\mathrm{p}=x_\mathrm{u}$, which minimizes path loss but ceases to be optimal once fading is introduced. We evaluate this strategy under fading, together with the placement proposed in \cite[Prop. 4]{tyrovolas_performance_2026}, which additionally accounts for in-\gls{WG} losses. Furthermore, while \glspl{PASS} inherently enable continuous \gls{PA} positioning, we compare this with a discrete \gls{PASS}, where ports are equally spaced at intervals of $0.1$ m along the \gls{WG}, and one can be activated at a time. All three approaches, alongside random and optimal positioning, are shown for discrete and continuous \glspl{PASS}. Here, $L{=}5$ m and $d_\mathrm{LS}{=}5$ m.

The continuous \gls{PASS} outperforms its discrete counterpart, with the performance gap most pronounced under optimal \gls{PA} placement. The finer spatial resolution afforded by a continuous \gls{PASS} enables the \gls{PA} to be positioned at the true optimum, whereas discrete positioning is constrained to a finite set of candidate locations. Nevertheless, discrete positioning still achieves strong performance, indicating that the proposed methods remain effective even under this constraint. Furthermore, positioning informed by the random look-based \gls{REM} outperforms nearest-to-\gls{UE} positioning, since the latter does not account for shadow fading and therefore fails to exploit spatial variations in the channel. Nearest-to-\gls{UE} positioning nonetheless performs reasonably well, as path loss remains the dominant contributor to the overall channel gain. Finally, incorporating \gls{WG} attenuation at values typical of those reported in the literature \cite{ding_flexible_2025} has a negligible effect on the performance of nearest-to-\gls{UE} positioning, suggesting that this loss mechanism is of secondary importance relative to positioning strategy.

\section{Conclusions and Future Work}

\subsection{Conclusion}

This paper provided the first analytical study of \glspl{PASS} in Ricean fading environments, focusing on a fundamental \gls{SU} system with spatially correlated shadowing. Accurate expressions for the distribution of the maximum achievable \gls{SNR} were obtained for both exponentially and smoothly correlated \gls{LSF}, as well as for the mean \gls{SNR} under exponentially correlated \gls{LSF}. Simulation results verified the accuracy of the analytical framework.

The results show that incorporating fading effects fundamentally alters PA position selection compared with purely LoS environments. To address this, a low-complexity two-stage approach was proposed, where \glspl{REM} were used to determine an initial \gls{PA} position based on the \gls{LSF} which was subsequently refined by sampling fast-fading realizations in the surrounding region. The proposed methods achieved near-optimal performance with robustness to localization errors. 

Overall, this work provides an analytical framework for \glspl{PASS} beyond purely \gls{LoS} channels and shows the effectiveness of \gls{REM}-based positioning in scattered environments.

\subsection{Topics for Future Work}
\label{sec:FW}
As the first study to consider PASSs in Ricean environments, this work opens several avenues for future investigation, two of which are outlined below.

\subsubsection{Position-Dependent $K$-Factor Modeling}
An important direction for future work is the extension of the Ricean $K$-factor to a position-dependent model. A position-dependent $K$-factor would offer a more realistic representation of the propagation environment, particularly in scenarios where obstacles or \gls{WG} geometry cause the strength of the \gls{LoS} path to vary across the \gls{WG}, and represents a valuable research topic.

\subsubsection{Multi-Interferer SIR Analysis}

While the methods in this work apply to the \gls{SIR} with a single interferer, the scenario involving multiple interferers is more complex. In the presence of multiple interferers, the \gls{SIR} can be expressed as $\mathrm{SIR}=\beta_U/\sum_{i=1}^{K_I}\beta_i$, where $K_I$ is the number of interferers. Using the $K_I=2$ case as an example,
\begin{align}
    &\mathrm{SIR}=\frac{\beta_U}{\beta_1+\beta_2}, \notag \\ &=\!\frac{d_{U}(x_\mathrm{p})^{-\gamma}10^{\sigma_\mathrm{SF}X_U(x_\mathrm{p})/10}}{d_{1}(x_\mathrm{p})^{-\gamma}10^{\sigma_\mathrm{SF}X_1(x_\mathrm{p})/10}\!+\!d_{2}(x_\mathrm{p})^{-\gamma}10^{\sigma_\mathrm{SF}X_2(x_\mathrm{p})/10}}\!.\!
\end{align}
\noindent This is not analytically tractable. A classic approximation is to say that $\beta_1+\beta_2\approx\beta_\mathrm{I}$, where $\beta_\mathrm{I}$ is an approximate lognormal representation of the sum of the interferers. This approximation is well-known. However, to use this approximation, $\beta_\mathrm{I}$ must be converted to an equivalent shadowing/distance formulation, e.g. $\beta_\mathrm{I}=Ad_\mathrm{I}(x_\mathrm{p})^{-\gamma}10^{\sigma_\mathrm{SF}X_\mathrm{I}(x_\mathrm{p})/10}$. This may be possible, and represents an important extension of the work.

\section*{Appendix A \\ Derivation of $g'(s,x)$}
The product rule can be used to differentiate $g(s,x)$, where
\begin{equation}
    g'(s,x)=\tfrac{10}{\sigma_\mathrm{SF}}\left[f(s)v'(s) + v(s)f'(s)\right], \label{eq:gdash_appen}
\end{equation}
and $v(s)=\sqrt{s+1}$, $v'(s)=\tfrac{1}{2\sqrt{s+1}}$, while
\begin{equation}
    f(s)=\log_{10}\!\left(\tfrac{\sigma^2x\,d\left(\Omega(s)\right)^\gamma}{E_sA\,|\beta_\mathrm{W}\left(\Omega(s)\right)|^2}\right).
\end{equation}
To derive $f'(s)$, we begin by changing the base of the logarithm and applying the logarithm product rule to get
\begin{equation}
    u(s)\!=\!\tfrac{1}{\log(10)}\!\!\left[\log\!\left(\!\!\tfrac{\sigma^2x}{E_sA}\!\!\right)\!\!-\!\log\!\left(|\beta_\mathrm{W}(\Omega(s)|^2\right)\!+\!\log\!\left(d(\Omega(s))^\gamma\right)\right]\!. \notag
\end{equation}
Taking the derivative of each term using the chain rule, 
\begin{equation}
    u'(s)=\tfrac{1}{2\alpha\log(10)(s+1)}\!\!\left[\gamma\tfrac{d'(\Omega(s))}{d(\Omega(s))}{-}\tfrac{|\beta_\mathrm{W}(\Omega(s))|^{2'}}{|\beta_\mathrm{W}(\Omega(s))|^2}\right]\!\!,
\end{equation}
where
\begin{equation}
    \tfrac{|\beta_\mathrm{W}(\Omega(s))|^{2 '}}{|\beta_\mathrm{W}(\Omega(s))|^2}\!=\!\tfrac{d}{d\Omega(s)}\!\left(\!-\tfrac{\kappa\log(10)}{10}\Omega(s)\!\right)\!=\! -\tfrac{\kappa\log(10)}{10},
\end{equation}
and
\begin{equation}
    d'(\Omega(s)) = \tfrac{\Omega(s)-x_\mathrm{u}}{d(\Omega(s))}.
\end{equation}
Thus, combining the product rule terms with \eqref{eq:gdash_appen} gives \eqref{eq:gdash}.

\section*{Appendix B \\ Integral over $u$ in $P_2$}
To solve the integral over $u$ in \eqref{eq:P2_doubleint}, we must first isolate the components involving $u$ and apply linearity, giving
\begin{multline}
    P_2\approx\!\! \int_0^{\mathrm{e}^{2\alpha L}\!-1}\!\!\!\frac{1}{2\pi\sqrt{s}}\!\!\left(\!\!\frac{g(s,x)}{s}\!-\!g'(s,x)\!\!\right)\!\exp\!\left(\!\!{-\frac{g^2(s,x)}{2s}}\!\!\right)\!I_1\,ds \\ - \int_0^{\mathrm{e}^{2\alpha L}-1}\!\frac{1}{2\pi s\sqrt{s}}\exp\!\left(\!{-\frac{g^2(s,x)}{2s}}\!\right)I_2\,ds, \label{eq:P2_ints}
\end{multline}
where
\begin{align}
    I_1&= \int_{-\infty}^{y(0,x)} \exp\left({-u^2\!\left(\frac{1}{2}+\frac{1}{2s}\right)+u\frac{g(s,x)}{s}}\right)du,\label{eq:intI1} \\ 
    I_2&= \int_{-\infty}^{y(0,x)} u\,\exp\left(\!-u^2\!\left(\frac{1}{2}+\frac{1}{2s}\right)\!+u\frac{g(s,x)}{s} \right) du,\label{eq:intI2}
\end{align}
are the integrals over $u$ to be solved. Letting $b_I=\frac{1}{2}+\frac{1}{2s}$ and $c_I=g(s,x)/s$ and completing the square, $I_1$ becomes
\begin{equation}
    I_1=\exp\left(\frac{c_I^2}{4b_I}\right)\int_{-\infty}^{y(0,x)}\exp\left(-b_I\left(u-\frac{c_I}{2b_I}\right)^2\right)du. \notag
\end{equation}
Making the substitution $t=u-\frac{c_I}{2b_I}$, $du=dt$,
\begin{equation}
    I_1=\exp\left(\frac{c_I^2}{4b_I}\right)\int_{-\infty}^{y(0,x)-c_I/2b_I}\exp\left(-b_It^2\right)dt. \notag
\end{equation}
The integral is now in the form of a normal \gls{CDF}. Therefore,
\begin{equation}
    I_1\!=\!\sqrt{\tfrac{\pi}{4b_I}}\exp\!\left(\!\tfrac{c_I^2}{4b_I}\!\right)\!\!\left(\!1\!+\!\mathrm{erf}\!\left(\!\!\sqrt{b_I}\!\left(y(0,x)\!-\!\tfrac{c_I}{2b_I}\!\right)\!\!\right)\!\!\right)\!. \label{eq:I1}
\end{equation}
The second term, $I_2$, has a form similar to $I_1$ with an additional factor of $u$. Again, expressing the integral in terms of $b_I$ and $c_I$ and making the same substitution for $t$ yields
\begin{align}
    I_2&=\exp\left(\frac{c_I^2}{4b_I}\right)\int_{-\infty}^{y(0,x)-c_I/2b_I}\left(t+\frac{c_I}{2b_I}\right)\exp\left(-b_It^2\right)dt, \notag \\
    &=\exp\left(\frac{c_I^2}{4b_I}\right)\bigg[\int_{-\infty}^{y(0,x)-c_I/2b_I}t\exp\left(-b_It^2\right)dt \notag \\&\qquad\qquad\qquad\quad+\frac{c_I}{2b_I}\int_{-\infty}^{y(0,x)-c_I/2b_I}\exp\left(-b_It^2\right)dt\bigg], \notag \\
    &=\exp\left(\tfrac{c_I^2}{4b_I}\right)\left[I_{21}+\tfrac{c_I}{2b_I}I_{22}\right]. \label{eq:I2}
\end{align}
Recognizing that $\frac{d}{dt}\big(\!-\!\frac{1}{2b_I}\exp\left(-b_It^2\right)\!\big)=t\exp\left(-b_It^2\right)$,
\begin{equation}
    I_{21}=-\tfrac{1}{2b_I}\exp\!\Big(\!-\!b_I(y(0,x)-\tfrac{c_I}{2b_I})^{2}\Big). \label{eq:I21}
\end{equation}
From the derivation of $I_1$,
\begin{equation}
    I_{22}=\sqrt{\tfrac{\pi}{4b_I}}\left(1+\mathrm{erf}\left(\sqrt{b_I}\left(y(0,x)-\tfrac{c_I}{2b_I}\right)\right)\right). \label{eq:I22}
\end{equation}
Combining \eqref{eq:I1}, \eqref{eq:I2}, \eqref{eq:I21},  and \eqref{eq:I22} with \eqref{eq:P2_ints} gives \eqref{eq:P2_approx}.

\section*{Appendix C \\ Integral over $x$ in $T_2$}
The expression for $T_2$ in \eqref{eq:T2int} can be written as 
\begin{multline}
    T_2\!=\!\int_0^{\mathrm{e}^{2\alpha L}\!-1}\!\!\!\!\int_0^\infty\!\!(B\log(x)\!+\!C)\exp\!\left(-(D\log(x)\!+\!E)^2\right) \\ \times (1+\mathrm{erf}(F\log(x)+G))dxds,
\end{multline}
where $B\!=\!\frac{10(s+1)-5s}{\sigma_\mathrm{SF}\!\sqrt{8\pi}\log(10)s(s+1)}$, $D\!=\!\frac{\sqrt{50}}{\sigma_\mathrm{SF}\!\log(10)}$, $E\!=\!\frac{\sqrt{50}\Psi(\Omega(s))}{\sigma_\mathrm{SF}\!\log(10)}$, $F\!\!=\!\!\frac{10(\sqrt{s+1}-1)}{\sigma_\mathrm{SF}\!\log(10)\!\sqrt{2s}}$, $G\!=\!\!\frac{10(\sqrt{s+1}\Psi(0)-\Psi(\Omega(s)))}{\sigma_\mathrm{SF}\!\log(10)\!\sqrt{2s}}$ and 
\begin{multline}
    C = \tfrac{5}{\sigma_\mathrm{SF}\sqrt{8\pi}\log(10)(s+1)}\bigg[\tfrac{2(s+1)-s}{s}\Psi(\Omega(s)) \\ -\tfrac{\kappa\log(10)}{10\alpha}-\tfrac{\gamma(\Omega(s)-x_\mathrm{u})}{\alpha d(\Omega(s))^2}\bigg]. 
\end{multline}
Letting $y=F\log(x)+G$ and $\frac{dx}{dy}=\frac{1}{F}\exp(\frac{y-G}{F})$ gives
\begin{equation}
    T_2\!=\!\int_0^{\mathrm{e}^{2\alpha L}\!-1}\!\frac{\exp\left(-G/F\right)}{F} I_{T_2}ds, \label{eq:T2appen}
\end{equation}
where
\begin{equation}
    I_{T_2}\!=\!\!\!\int_{-\infty}^\infty\!\!\!\left(\!\tilde{B}y\!+\!\tilde{C}\!\right)\!\exp\!\Big(\!\!-\!\!\Big(\!\!\left(\!\tilde{D}y\!+\!\tilde{E}\!\right)^{\!2}\!\!\!-\!\frac{y}{F}\!\Big)\!\!\Big)\!(1\!+\!\mathrm{erf}(y)\!)dy,
\end{equation}
and $\tilde{B}=\frac{B}{F}$, $\tilde{C}=C-\frac{BG}{F}$, $\tilde{D}=\frac{D}{F}$ and $\tilde{E}=E-\frac{DG}{F}$. By completing the square, the argument of the exponential function is $(\tilde{D}y+\tilde{E})^2-\frac{y}{F}=(\tilde{D}y+H)^2+J$, where $H=\tilde{E}-\frac{1}{2\tilde{D}F}$ and $J=\frac{\tilde{E}}{\tilde{D}F}-\frac{1}{4\tilde{D}^2F^2}$. Expanding brackets,
\begin{align}
    I_{T_2}= & \mathrm{e}^{-J}\!\bigg[\!\int_{-\infty}^\infty(\tilde{B}y\!+\!\tilde{C})\!\exp(\!-(\tilde{D}y\!+\!H)^2)\mathrm{erf}(y) dy\notag \\ & \quad+ \int_{-\infty}^\infty(\tilde{B}y\!+\!\tilde{C})\exp(-(\tilde{D}y+H)^{2})dy \bigg]\!, \notag \\
    =&\mathrm{e}^{-J}\left[I_{T_2,1} + I_{T_2,2}\right]. \label{eq:IT2}
\end{align}
We can write $(\tilde{B}y+\tilde{C})=\frac{\tilde{B}}{\tilde{D}}(\tilde{D}y+H)+\tilde{C}-\frac{\tilde{B}H}{\tilde{D}}$, giving
\begin{align}
    I_{T_2,1}=&\tfrac{\tilde{B}}{\tilde{D}}\int^\infty_{-\infty}(\tilde{D}y+H)\exp(-(\tilde{D}y+H)^2)\mathrm{erf}(y)dy \notag \\
    &+ \Big(\tilde{C}-\tfrac{\tilde{B}H}{\tilde{D}}\Big)\int^\infty_{-\infty}\exp(-(\tilde{D}y+H)^2)\mathrm{erf}(y)dy, \notag \\
    =&\tfrac{\tilde{B}}{\tilde{D}}I_{T_2,1,1}+\Big(\tilde{C}-\tfrac{\tilde{B}H}{\tilde{D}}\Big)I_{T_2,1,2}. \label{eq:T21}
\end{align}
Integration by parts can be used to solve $I_{T_2,1,1}$, such that
\begin{align}
    I_{T_2,1,1} &= \bigg[-\frac{\mathrm{erf}(y)\exp(-(\tilde{D}y+H)^2)}{2\tilde{D}}\bigg]^\infty_{-\infty} \notag \\ &\qquad\,\, +\frac{1}{2\tilde{D}}\int_{-\infty}^\infty\!\!\left(\!\frac{d}{dy}\mathrm{erf}(y)\!\!\right)\!\exp(-(\tilde{D}y+H)^2)dy, \notag \\
    &=  0 + \frac{1}{\sqrt{\pi}\tilde{D}}\int_{-\infty}^\infty\exp(-(\tilde{D}y+H)^2-y^2)dy.
\end{align}
Therefore, by completing the square such that $(\tilde{D}y+H)^2+y^2\!=\!(\tilde{D}^2+1)\!\left(\!y+\frac{\tilde{D}H}{\tilde{D}^2\!+\!1}\!\right)^{\!\!2}\!\!+\!\frac{H^2}{\tilde{D}^2+1}$, the final result for $I_{T_2,1,1}$ is
\begin{equation}
    I_{T_2,1,1}=\tfrac{1}{\tilde{D}\sqrt{\tilde{D}^2+1}}\exp\Big(-\tfrac{H^2}{\tilde{D}^2+1}\Big). \label{eq:T211}
\end{equation}
The result in \cite[Eq. (13)]{ng_table_1969} gives $I_{T_2,1,2}$ as
\begin{equation}
    I_{T_2,1,2}=-\tfrac{\sqrt{\pi}}{\tilde{D}}\mathrm{erf}\Big(\tfrac{H}{\sqrt{\tilde{D}^2+1}}\Big). \label{eq:T212}
\end{equation}
To solve $I_{T_2,2}$, the substitution $t\!=\!\!\tilde{D}y+\!H$ can be made, giving
\begin{equation}
    I_{T_2,2}\!=\!\!\Big(\!\tfrac{\tilde{C}}{\tilde{D}}\!-\!\tfrac{\tilde{B}H}{\tilde{D}^2}\!\Big)\!\!\int_{-\infty}^\infty\!\!\exp\left(-t^2\right)dt=\!\!\Big(\!\tfrac{\tilde{C}}{\tilde{D}}\!-\!\tfrac{\tilde{B}H}{\tilde{D}^2}\!\Big)\!\sqrt{\pi}. \label{eq:T22}
\end{equation}
Combining 
\eqref{eq:IT2},
\eqref{eq:T21}, \eqref{eq:T211}, \eqref{eq:T212} and \eqref{eq:T22} with \eqref{eq:T2appen} gives \eqref{eq:T2}.

\section*{Appendix D \\ Integral over $x$ in $T_3$}
The expression for $T_3$ in \eqref{eq:T3int} can be written as
\begin{align}
    T_3= \int_0^{\mathrm{e}^{2\alpha L}\!-1}I_{T_3,1}\,ds - \int_0^{\mathrm{e}^{2\alpha L}\!-1}I_{T_3,2}\,ds, \label{eq:T3appen}
\end{align}
where
\begin{multline}
    I_{T_3,1} = \int_0^\infty\!\!\frac{1}{2\sqrt{2\pi(s+1)}}\frac{g(s,x)}{s(s+1)}\mathrm{exp}\bigg(\!\!\!-\frac{g^2(s,x)}{2(s+1)}\bigg) \\ \,\times\!\!\bigg(\!1\!+\!\mathrm{erf}\bigg(\!\sqrt{\!\frac{s\!+\!1}{2s}}y(0,x)\!-\!\frac{g(s,x)}{\sqrt{2s(s\!+\!1)}}\bigg)\!\!\bigg)dx,
\end{multline}
and
\begin{multline}
    I_{T_3,2}\!=\!\! \int_0^\infty\!\frac{1}{2\pi\sqrt{s}(s+1)}\exp\!\bigg(\frac{y(0,x)g(s,x)}{s} \\ -\frac{g^2(s,x)}{2s}-\frac{(s+1)y(0,x)^2}{2s}\bigg)dx.
\end{multline}
Note that $I_{T_3,1}$ has the same form as $T_2$ and thus can be written
\begin{multline}
    I_{T_3,1}\!=\!\!\int_0^\infty\!\!(M\log(x)\!+\!Q)\exp\!\left(-(D\log(x)\!+\!E)^2\right) \\ \times (1+\mathrm{erf}(F\log(x)+G))dx,
\end{multline}
where $M=\frac{5}{\sqrt{2\pi}s(s+1)\sigma_\mathrm{SF}\log(10)}$, $Q=\frac{5\Psi(\Omega(s))}{\sqrt{2\pi}s(s+1)\sigma_\mathrm{SF}\log(10)}$, and $D$, $E$, $F$ and $G$ are as defined in Appendix C. Following the same procedure as in Appendix C,
\begin{multline}
    I_{T_3,1}\!=\tfrac{\exp(-(\tfrac{G}{F}+J))}{F\tilde{{D}}}\bigg[\tfrac{\tilde{M}\exp\big(-\frac{H^2}{\tilde{D}^2+1}\big)}{\tilde{D}\sqrt{\tilde{D}^2+1}} \\ +\sqrt{\pi}\bigg(\!\!\tilde{Q}\!-\!\tfrac{\tilde{M}H}{\tilde{D}}\bigg)\bigg(\!\!1-\mathrm{erf}\bigg(\!\tfrac{H}{\sqrt{\tilde{D}^2+1}}\bigg)\!\!\bigg)\!\bigg], \label{eq:IT31}
\end{multline}
where $\tilde{M}=\frac{M}{F}$, $\tilde{Q}=Q-\frac{MG}{F}$, and $\tilde{D}$, $H$ and $J$ are as defined in Appendix C. Letting $y=\log(x)$, $I_{T_3,2}$ becomes
\begin{equation}
    I_{T_3,2}=\tfrac{\exp(c)}{2\pi\sqrt{s}(s\!+\!1)}\!\int^\infty_{-\infty}\!\exp\left(ay^2\!+\!(b+1)y\right)dy,
\end{equation}
where $a\!=\!\frac{100\sqrt{s+1}(1-\sqrt{s+1})}{s\sigma_\mathrm{SF}^2\log(10)^2}$, $b\!=\!\frac{100\sqrt{s+1}(1-\sqrt{s+1})}{s\sigma_\mathrm{SF}^2\log(10)^2}(\Psi(\Omega(s))\!+\!\Psi(0))$ and $c\!=\!-\frac{50\sqrt{s+1}}{s\sigma_\mathrm{SF}^2\log(10)^2}(\sqrt{s+1}(\Psi(\Omega(s))^2+\Psi(0)^2)-2\Psi(\Omega(s))\Psi(0))$. Applying the result in \cite[Eq. (3.323.2)]{gradshteyn_table_2007},
\begin{equation}
    I_{T_3,2}=\tfrac{1}{2(s+1)\sqrt{-a \pi s}}\exp\Big(c-\tfrac{(b+1)^2}{4a}\Big). \label{eq:IT32}
\end{equation}
Therefore, combining \eqref{eq:IT31} and \eqref{eq:IT32} with \eqref{eq:T3appen} gives \eqref{eq:T3}.

\bibliographystyle{IEEEtran}
\bibliography{references}

\begin{IEEEbiography}[{\includegraphics[width=1in,height=1.25in,clip,keepaspectratio]{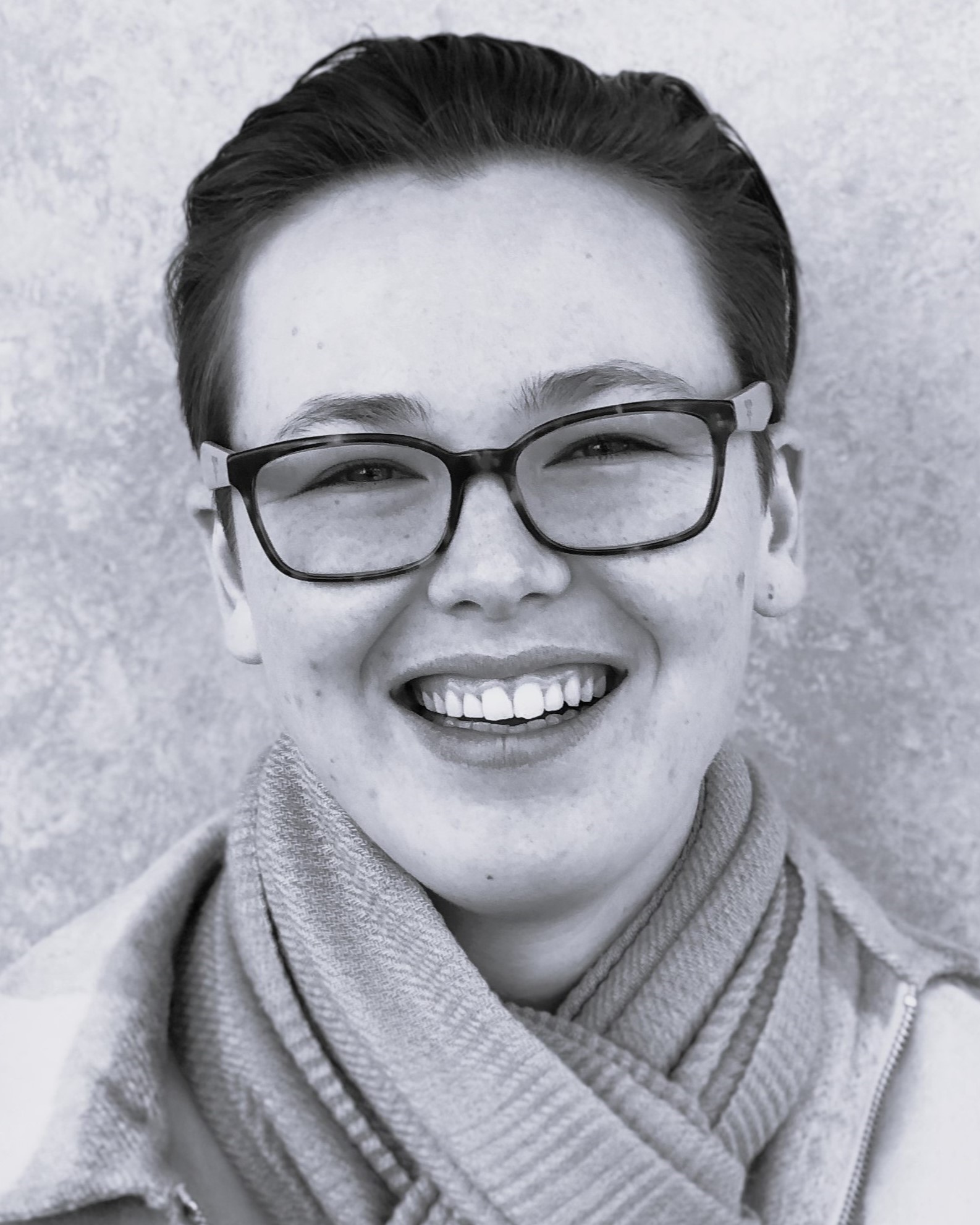}}]%
{Amy S. Inwood} (Member, IEEE) received the B.E (Hons.) and Ph.D. degrees in electrical and electronic engineering from Te Whare Wānanga o Waitaha $|$ University of Canterbury (UC), NZ, in 2021 and 2024, respectively. She is now a Research Fellow at the Centre for Wireless Innovation, Queen's University Belfast, U.K. Her research interests include the statistical analysis of wireless communications systems, MIMO, reconfigurable intelligent surfaces, fluid antenna systems, pinching antenna systems and other next-generation antenna technologies. She is currently an Editor for Elsevier Physical Communication and an Associate Editor for \textsc{IEEE Wireless Communications Letters}.
\end{IEEEbiography}
\begin{IEEEbiography}[{\includegraphics[width=1in,height=1.25in,clip,keepaspectratio]{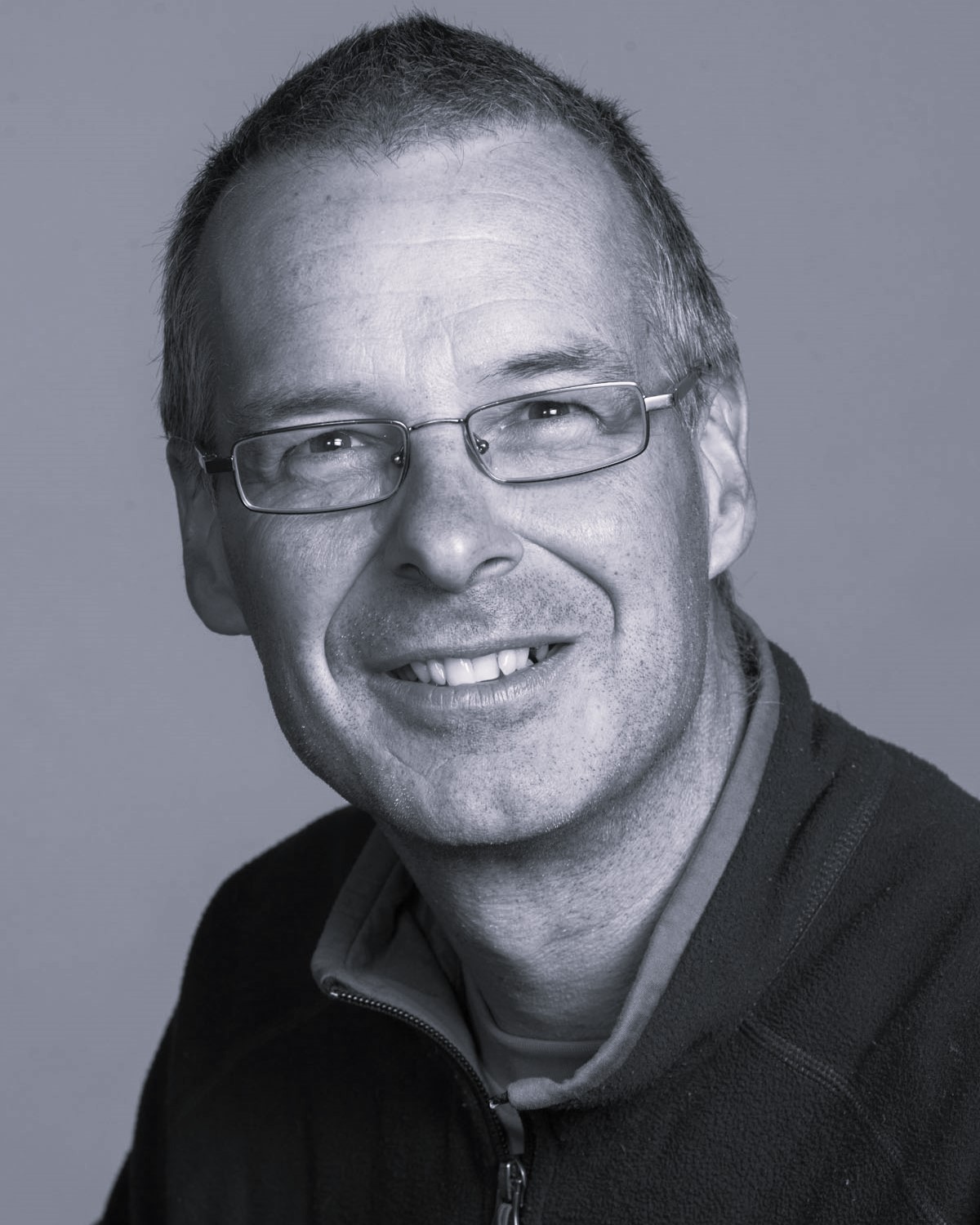}}]%
{Peter J. Smith} (Fellow, IEEE) received the B.Sc degree in Mathematics and the Ph.D degree in Statistics from the University of London, London, U.K., in 1983 and 1988, respectively. In 2015 he joined Victoria University of Wellington as Professor of Statistics. He is an Honorary Professor in the School of Electronics, Electrical Engineering and Computer Science, Queens University Belfast. His research interests include the statistical aspects of design, antenna arrays, MIMO, cognitive radio, mmWave systems, reconfigurable intelligent surfaces, fluid antenna systems and integrated sensing and communications.
\end{IEEEbiography}

\begin{IEEEbiography}[{\includegraphics[width=1in,height=1.25in,clip,keepaspectratio]{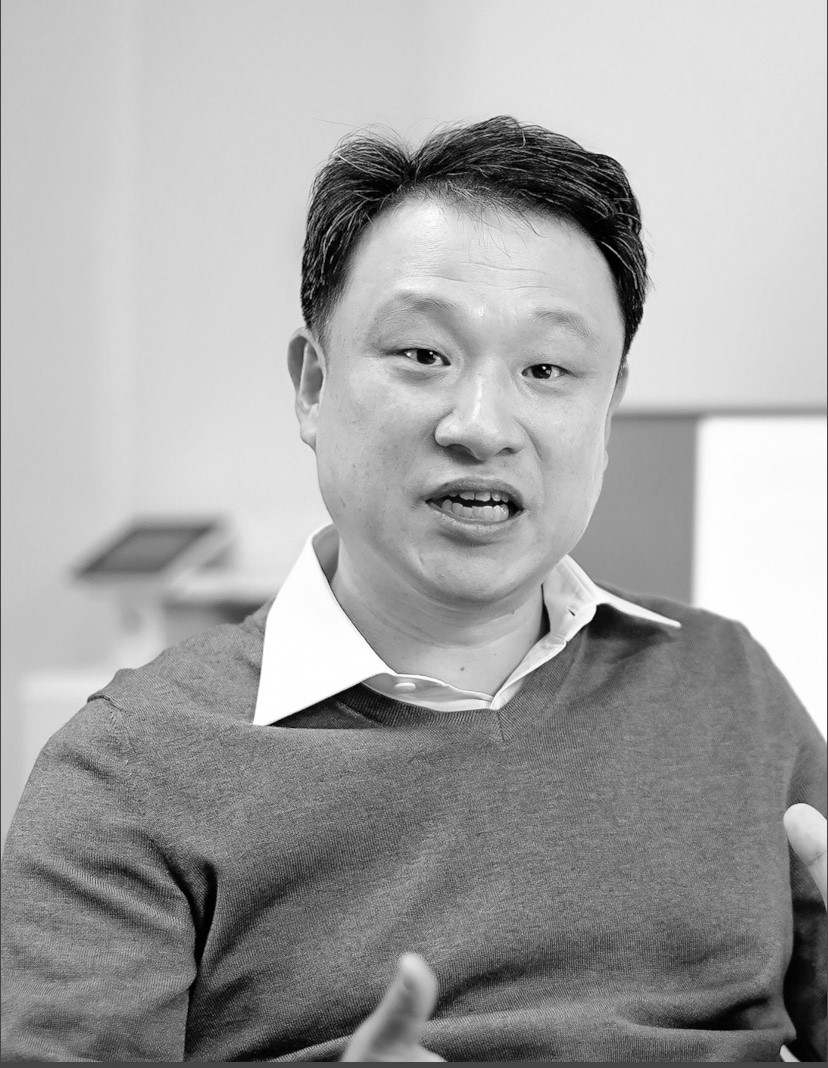}}]%
{Hyundong Shin} (Fellow, IEEE)  received the B.S. degree in electronics engineering from Kyung Hee University (KHU), Yongin-si, South Korea, in 1999, and the M.S. and Ph.D. degrees in electrical engineering from Seoul National University, Seoul, South Korea, in 2001 and 2004, respectively. During his post-doctoral research with Massachusetts Institute of Technology (MIT) from 2004 to 2006, he was with the Laboratory for Information Decision Systems (LIDS). In 2006, he joined KHU, where
he is currently a Professor with the Department of
Electronic Engineering. His research interests include quantum information science, wireless communication, and machine intelligence. He received the IEEE Communications Society’s Guglielmo Marconi Prize Paper Award and the William R. Bennett Prize Paper Award. He served as the Publicity Co-Chair for IEEE PIMRC and the Technical Program Co-Chair for IEEE
WCNC and IEEE GLOBECOM. He was an Editor of \textsc{IEEE Transactions on Wireless Communications} and \textsc{IEEE Communications Letters}.
\end{IEEEbiography}
\begin{IEEEbiography}[{\includegraphics[width=1in,height=1.55in,clip,keepaspectratio]{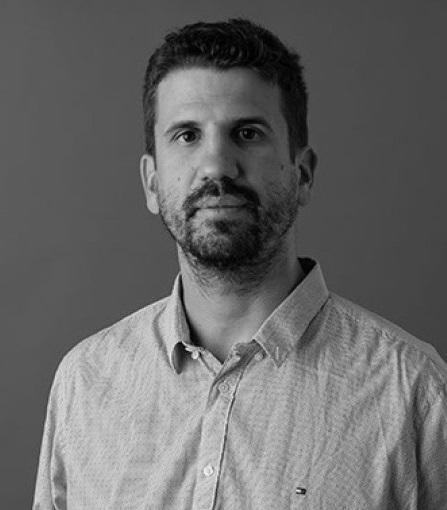}}]
{Michail Matthaiou} (Fellow, IEEE) obtained his Ph.D. degree from the University of Edinburgh, U.K. in 2008. He is currently a Professor of Communications Engineering and Signal Processing and Deputy Director of the Centre for Wireless Innovation (CWI) at Queen’s University Belfast, U.K. He is also an Eminent Scholar at the Kyung Hee University, Republic of Korea. He has held research/faculty positions at Munich University of Technology (TUM), Germany and Chalmers University of Technology, Sweden. His research interests span signal processing for wireless communications, beyond massive MIMO, reflecting intelligent surfaces, mm-wave/THz systems and AI-empowered communications. Dr. Matthaiou and his coauthors received the IEEE Communications Society (ComSoc) Leonard G. Abraham Prize in 2017. He currently holds the ERC Consolidator Grant BEATRICE (2021-2026) focused on the interface between information and electromagnetic theories. He has received the prestigious 2023 Argo Network Innovation Award, the 2019 EURASIP Early Career Award and the 2018/2019 Royal Academy of Engineering/The Leverhulme Trust Senior Research Fellowship. His team was also the Grand Winner of the 2019 Mobile World Congress Challenge. He was the recipient of the 2011 IEEE ComSoc Best Young Researcher Award for the Europe, Middle East and Africa Region and a co-recipient of the 2006 IEEE Communications Chapter Project Prize for the best M.Sc. dissertation in the area of communications. He has co-authored papers that received best paper awards at the 2018 IEEE WCSP and 2014 IEEE ICC. In 2014, he received the Research Fund for International Young Scientists from the National Natural Science Foundation of China. He is currently the Editor-in-Chief of Elsevier Physical Communication, a Senior Editor for \textsc{IEEE Wireless Communications Letters}, and  an Area Editor for \textsc{IEEE Transactions on Communications} and Editor-in-Large for \textsc{IEEE Open Journal of the Communications Society}. He is an IEEE and AAIA Fellow.
\end{IEEEbiography}

\end{document}